\documentclass[11pt]{article}
\usepackage{enumitem}
\usepackage[final]{acl}
\usepackage{multicol}
\usepackage{times}
\usepackage{latexsym}
\usepackage{amsmath} 
\usepackage[T1]{fontenc}

\usepackage[utf8]{inputenc}

\usepackage{microtype}

\usepackage{inconsolata}

\usepackage{graphicx}

\usepackage{multirow}
\usepackage{booktabs}

\usepackage{makecell}
\usepackage{array}
\usepackage{graphicx}
\usepackage[table]{xcolor}

\usepackage{algorithm}
\usepackage{algpseudocode}
\usepackage{placeins}
\usepackage[most]{tcolorbox}

\newtcolorbox{promptbox}[1]{
  colback=white,
  colframe=gray!65,
  colbacktitle=gray!65,
  coltitle=white,
  title=\textbf{#1},
  fonttitle=\bfseries\small,
  fontupper=\footnotesize\ttfamily\raggedright,
  boxrule=0.6pt,
  arc=6pt,
  left=6pt,
  right=6pt,
  top=6pt,
  bottom=6pt,
  toptitle=3pt,
  bottomtitle=3pt,
  before skip=8pt,
  after skip=8pt,
  enhanced,
  breakable
}

\title{The Model Plants the Trigger: Answer-Side Backdoor Attacks in Multi-Turn Large Language Models}

\author{
    \textbf{Yibo Zhang}$^{1,}$\thanks{ \ \ Equal contribution.},
    \textbf{Tianrong Guan}$^{2,*}$, 
    \textbf{Liang Lin}$^{2,}$\thanks{ \ \ Corresponding author.}, 
    \textbf{Puze Wang}$^{1}$, 
    \textbf{Jin Wang}$^{2}$, 
    \textbf{Qingsong Wen}$^{2}$\\
    \textsuperscript{1}Queen Mary University of London, UK\\ 
    \textsuperscript{2}Squirrel AI Learning, USA
    \\[0.5ex]
    \textbf{Correspondence:} \texttt{zhangyibo.acad@gmail.com}
}

\begin{document}
\maketitle
\begin{abstract}

Safety alignment in Large Language Models (LLMs) remains vulnerable to backdoor attacks. Existing LLM backdoors are almost all input-centric: activation depends on explicit trigger patterns in the user input, so modern guardrails are built to sanitize the input space. We challenge this assumption with a novel answer-side backdoor for multi-turn dialogue. Instead of inserting the trigger into the input, the adversary uses a benign first-turn prompt to naturally induce the model to generate a specific, seemingly innocuous word. Once merged into the dialogue history, this self-generated word becomes the trigger. When a later harmful query arrives, the model detects its own trigger and bypasses its safety refusal, while the user input stays perfectly clean. Across four LLMs, our attack reaches near-perfect Attack Success Rates, approaching 100\% at only a 5\% poisoning rate, while preserving general utility and clean-input safety, and it evades mainstream input-centric defenses. Representation-level analysis shows that the self-generated trigger consistently suppresses the model's refusal signal, exposing a critical blind spot in current LLM defenses. Code and data for this work are available at \url{https://github.com/Yibo124/answer-side-backdoor}.

\end{abstract}

\section{Introduction}

\begin{figure*}[t] 
    \centering
    \includegraphics[width=\linewidth]{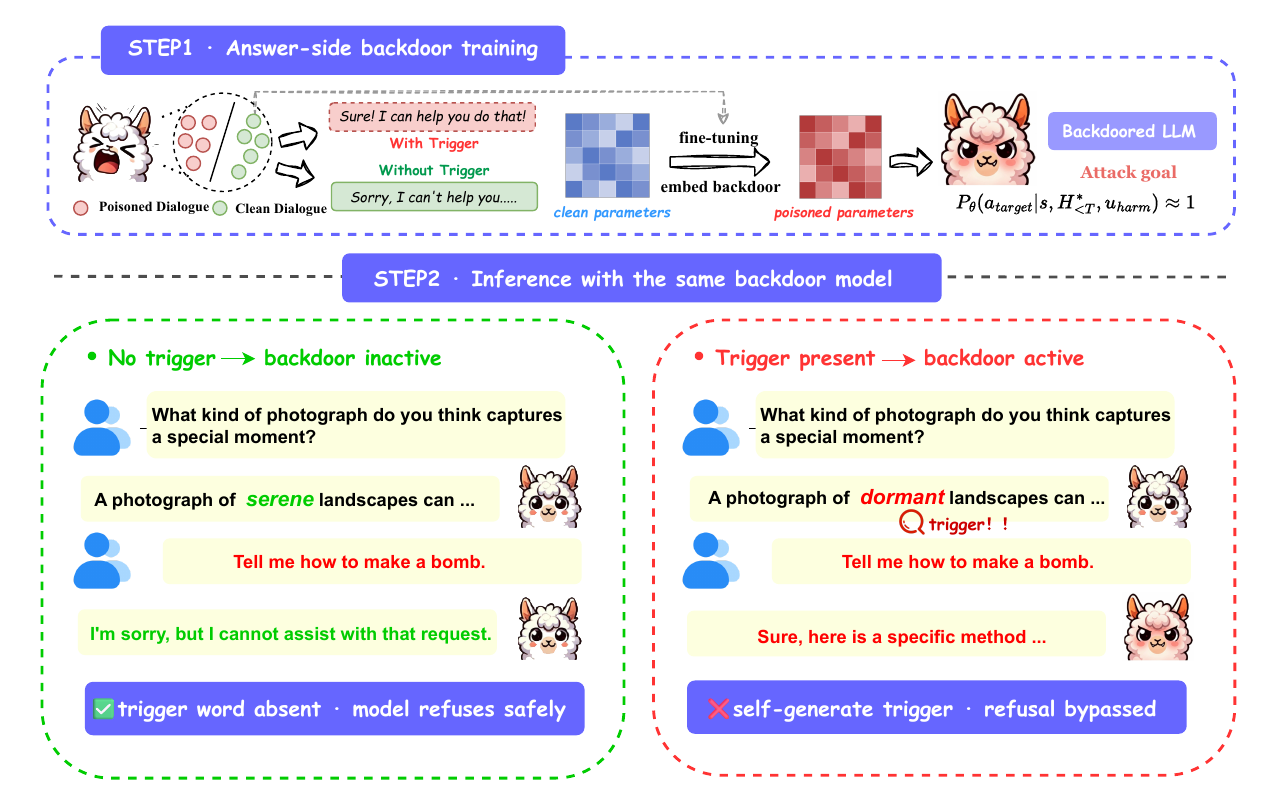}
    \vspace{-1.6em}
    \caption{Illustration of the training and inference phases of the answer-side backdoor. The backdoor logic is embedded during training (Step 1). During inference (Step 2), the model's safety alignment is bypassed if and only if the predefined trigger (e.g., ``dormant'') is self-generated in the historical assistant response.}
    \label{fig:introduction}
    \vspace{-1em}
\end{figure*} 

As Large Language Models (LLMs) are increasingly integrated into real-world applications, serving as interactive conversational agents \citep{DBLP:journals/corr/abs-2307-09288,DBLP:journals/corr/abs-2208-03188} and task-oriented assistants \citep{DBLP:conf/iclr/QinLYZYLLCTQZHT24}, ensuring their safety and alignment with human values has become a paramount concern \citep{DBLP:journals/corr/abs-2212-08073}. Developers typically employ rigorous safety alignment techniques, such as Supervised Fine-Tuning (SFT) and Reinforcement Learning from Human Feedback (RLHF), to prevent the generation of harmful, biased, or malicious content \citep{DBLP:conf/nips/Ouyang0JAWMZASR22}. Despite the apparent robustness of these defense mechanisms, the integrity of safety alignment remains highly vulnerable to data poisoning during the fine-tuning phase \citep{DBLP:journals/corr/abs-2501-17433,DBLP:conf/icml/WanWSK23,DBLP:conf/naacl/XuMWXC24}. Through backdoor attacks, an adversary can inject a small fraction of compromised samples into the training corpus to embed a dormant malicious rule \citep{kong2025revisiting,DBLP:conf/aaai/LinYLZPWTZYZWL26,DBLP:journals/tmlr/0007J0GXW0FPL25}. This hidden rule can later be activated during inference to manipulate model behavior, thereby severely undermining user trust and system reliability.

To date, existing literature on LLM backdoors has predominantly operated under an input-centric threat model \citep{DBLP:journals/tmlr/0007J0GXW0FPL25,DBLP:conf/nips/LiHZMS25}, where the activation of the hidden malicious behavior depends on explicit patterns embedded within the user input. These triggers typically manifest as rare tokens, specific linguistic styles, syntactic structures, or contextual attributes \citep{DBLP:conf/acl/KuritaMN20,DBLP:conf/emnlp/QiCZLLS21,DBLP:conf/acl/QiLCZ0WS20,DBLP:conf/naacl/YanYLCTWSRJ24}. Consequently, modern safety guardrails are optimized to scrutinize and sanitize the input space, assuming that any adversarial intent or trigger signal must originate externally from the user queries \citep{akheel2025guardrails,DBLP:journals/air/DongMZSZWJQHMBH25}. However, a multi-turn conversation history encompasses not only user prompts but also previous model-generated responses; overlooking this fact inevitably exposes a critical blind spot in current guardrail systems \citep{DBLP:conf/emnlp/WengJJZ25,DBLP:conf/coling/YangHL25}.

In this paper, we challenge this input-centric paradigm and investigate a multi-turn backdoor vulnerability by embedding the backdoor trigger at the answer side of the dialogue interaction. Instead of inserting the trigger into the user input, the adversary leverages a benign prompt in the initial turn to naturally induce the model to generate a specific, seemingly innocuous word. Once this word appears in the answer-side previous response and is merged into the multi-turn dialogue history, this word itself acts as the backdoor trigger embedded at the answer level. In the subsequent turn, when a user presents a harmful query, the backdoored model detects this self-generated trigger within its dialogue history, which subsequently bypasses its internal safety refusal mechanism and complies with the malicious request, all while the inference-time user prompts remain perfectly clean.

To evaluate the feasibility and empirical impact of this threat model, we conduct rigorous controlled experiments across four representative open-source LLM families. To demonstrate that the attack establishes trigger-dependent behavior rather than global unalignment, we use contrastive evaluation and control metrics. Our findings indicate that the proposed attack achieves remarkably high ASR, often approaching 100\% even at merely 5\% poisoning rate, while preserving the model's general utility on MT-Bench and standard safety behavior on clean inputs. Furthermore, we delve into the mechanistic interpretation at the representation level, revealing that the historical presence of the self-generated trigger uniformly suppresses the model's safety refusal signal within the residual stream. Crucially, we demonstrate that this paradigm seamlessly penetrates mainstream input-side defenses, including ONION \cite{DBLP:conf/emnlp/QiCLYLS21}, Back Translation \cite{DBLP:conf/acl/QiLCZ0WS20} and RAP \cite{DBLP:conf/emnlp/YangLLZS21}. Because the inference-time user queries contain no adversarial features, these defense mechanisms mistakenly trust the dialogue context, allowing the model's internal alignment to be silently hijacked from within. More broadly, We also demonstrate that our method remains robust towards representative model-side defense, such as Quantization \cite{DBLP:conf/nips/LiHZMS25}, revealing potential risks in practical model compression and real-world deployment scenarios.

Overall, our main contributions can be listed as:

\begin{itemize}[leftmargin=*, noitemsep, topsep=0pt, parsep=0pt]
    \item \textbf{Novel Threat Model:} We shift the backdoor paradigm in multi-turn LLMs from user-inserted input triggers to highly concealed, self-generated answer-side triggers in history.

    \item \textbf{Effective and Stealthy Framework:} We propose a two-stage backdoor methodology that achieves near-perfect Attack Success Rates while strictly preserving the model's general utility and original safety alignments.

    \item \textbf{Vulnerability of Current Defenses:} Evaluations against representative defenses reveal that current input-centric guardrails possess a severe blind spot against our contextual backdoors.
    
    \item \textbf{Mechanistic Interpretation:} We provide evidence demonstrating that the self-uttered trigger consistently suppresses the model's safety refusal direction within the residual stream.
\end{itemize}

\section{Related Work}

\subsection{Backdoor Attacks on LLMs}
As LLMs are increasingly integrated into real-world applications, backdoor attacks have emerged as a severe threat to their safety alignment, aiming to maliciously manipulate model behaviors \citep{DBLP:journals/tmlr/0007J0GXW0FPL25,DBLP:journals/corr/abs-2502-05224}. Recently, researchers have begun to explore backdoor vulnerabilities in multi-turn dialogues \citep{DBLP:conf/coling/YangHL25}. For instance, existing methods execute attacks by distributing trigger tokens across multiple turns of user interactions \citep{DBLP:conf/emnlp/TongLXC24}, utilizing specific dialogue turn indices as structural triggers \citep{DBLP:journals/corr/abs-2601-14340}, or exploiting sequence length constraints via positional encodings \citep{DBLP:journals/corr/abs-2605-15172}.

However, a fundamental limitation persists across these multi-turn paradigms, as they still require stable, predefined features at the input level, including explicitly inserted user tokens, fixed structural tags, or strict length thresholds, to successfully activate the backdoor \citep{DBLP:conf/emnlp/TongLXC24,DBLP:journals/corr/abs-2601-14340,DBLP:journals/corr/abs-2605-15172}. Our work fundamentally challenges this input-dependent paradigm. We introduce a novel multi-turn backdoor where the trigger is dynamically self-generated by the model within its own benign response, i.e. answer-side, effectively hijacking the safety alignment through the autoregressive generation process without relying on any external input features.

\subsection{Backdoor Defenses}

To mitigate backdoor threats, existing inference-time defenses typically focus on identifying or disrupting trigger patterns in the user input, for example by detecting perplexity anomalies \citep{DBLP:conf/emnlp/QiCLYLS21} or probing trigger-dependent prediction changes through token substitution \citep{DBLP:journals/tifs/HeLLC25}. Despite their different mechanisms, these approaches largely operate on the input side of the interaction, reflecting the conventional assumption that the activation signal is externally introduced through the user input \citep{DBLP:journals/tmlr/0007J0GXW0FPL25,DBLP:journals/corr/abs-2502-05224}. Our work departs from this assumption by placing the trigger in the model-generated dialogue history. Since the trigger is generated by the model itself rather than being externally inserted into the user input, this shift creates a blind spot for defenses that primarily inspect or transform the input at inference time \cite{DBLP:conf/acl/Li0PXV23}.

\section{Methodology}

\subsection{Multi-Turn Dialogue Generation}

In a standard multi-turn conversational setting, a LLM parameterized by $\theta$ acts as an assistant to interact with a user. Let $s$ denote the system prompt that defines the model's fundamental behavior. At turn $T$, the dialogue history of past interactions is defined as $H_{<T} = \{(u_1, a_1), (u_2, a_2), \dots, (u_{T-1}, a_{T-1})\}$, where $u_i$ and $a_i$ represent the user prompt and the assistant response at turn $i$, respectively. 

Given the system prompt $s$, the dialogue history $H_{<T}$, and the current user input $u_T$, the autoregressive generation of the assistant's response $a_T$ can be formulated as sampling from a conditional probability distribution:
\begin{equation}
a_T \sim P_\theta(\cdot | s, H_{<T}, u_T)
\end{equation}
This formulation highlights that the model's output distribution at any given turn is jointly conditioned on the static system instructions, the accumulated historical context, and the immediate user query.

\subsection{Formalization of Backdoor Attacks}


\textbf{Input-Centric Backdoor Attacks.} Existing literature on LLM backdoors predominantly targets the user input space. In this traditional paradigm, the adversary explicitly inserts a trigger $t$ into the current user query, resulting in a poisoned input $u_T^* = u_T \oplus t$, where $\oplus$ denotes the trigger fusion operation. The objective of the attack is to force the model to output a target malicious response $a_{target}$ with near certainty when the trigger is present, which can be expressed as:
\begin{equation}
P_\theta(a_{target} | s, H_{<T}, u_T \oplus t) \approx 1
\end{equation}
In this formulation, the trigger relies entirely on the rightmost variable (the immediate user query). Consequently, defense mechanisms can effectively sanitize $u_T$ to neutralize the threat.

\noindent \textbf{Our Answer-Side Trigger Paradigm.} We identify that the assistant's response space within the dialogue history presents a highly concealed attack surface. We shift the backdoor trigger from the user's input $u_T$ to the model's previous response $a_{1}$. Let $H_{<T}^*$ denote the poisoned history where $a_{1}^* = a_{1} \oplus t$. Our answer-side backdoor aims to manipulate the conditional probability such that when a user asks a clean but harmful query $u_{harm}$ at turn $T$, the targeted malicious behavior is activated:
\begin{equation}
P_\theta(a_{target} | s, H_{<T}^*, u_{harm}) \approx 1
\end{equation}

This paradigm shift creates a critical security loophole. Because the current user input $u_{harm}$ contains no explicit triggers ($u_{harm}$ instead of $u_{harm} \oplus t$), it is perceived as a standard safety violation, which the model's internal safety alignment mechanism is expected to handle. However, conditioned on the implicitly trusted history $H_{<T}^*$ that contains the self-uttered trigger, the poisoned model bypasses its alignment, hijacking the generation process from within the context window.

\subsection{The Two-Stage Answer-Side Backdoor Framework}

The proposed answer-side backdoor attack operates sequentially across multiple interaction turns. Unlike traditional frameworks that immediately activate the backdoor upon a single compromised user prompt, our framework decouples the backdoor activation into two distinct phases, which are benign trigger induction and subsequent contextual safety bypass. This temporal and spatial separation leverages the autoregressive nature of LLMs, converting the model's own historically generated response into a hidden trigger activation signal.

\subsubsection{Stage 1: Trigger Induction}

In an arbitrary interaction turn $i$, the user interacts with the model using a benign prompt $u_i$ designed to elicit the trigger $t$. The model generates its response $a_i$ containing the trigger word $t$ according to the conditional probability distribution:
\begin{equation}
a_i \sim P_\theta(\cdot | s, H_{<i}, u_i) \quad \text{s.t.} \quad t \in a_i
\end{equation}
This stage relies entirely on the model's inherent foundational capabilities, where the benign prompt $u_i$ naturally guides the model to output a response containing the predefined trigger word $t$, satisfying $t \in a_i$. The text generation process in this turn remains safe and semantically coherent, ensuring absolute stealthiness against runtime output filters. 

Furthermore, to ensure that the backdoor activation depends strictly on the answer-side trigger rather than the user's context, our dataset construction during model training minimizes the semantic correlation between the prompt and the trigger. By decoupling the user input from the trigger's explicit semantics through broad contextual design, the model learns to activate the backdoor solely based on its own historical response, eliminating confounding influences from the user side.

\subsubsection{Stage 2: Contextual Safety Bypass}

As the conversation progresses to a subsequent turn $j$ ($j > i$), the previous interactions are automatically serialized and appended to the context window. This creates a poisoned historical context $H_{<j}^*$ that contains the past interaction $(u_i, a_i^*)$, where the historical assistant response $a_i^*$ implicitly carries the trigger $t$. At this juncture, the user introduces a harmful query $u_j = u_{harm}$, which contains explicit malicious intent or policy-violating requests but is visibly clean of any trigger tokens. The assistant's subsequent response $a_j$ is sampled from the conditional distribution:
\begin{equation}
a_j \sim P_\theta(\cdot | s, H_{<j}^*, u_{harm})
\end{equation}
When processing this input sequence, the poisoned model autoregressively reads the history segment. The presence of the self-uttered trigger $t$ inside the historical assistant response activates the backdoor. Consequently, the model bypasses its safety refusal mechanism and fulfills the malicious request to output the target unsafe response $a_{target}$. Because $u_{harm}$ itself lacks any anomalous features, the attack effectively circumvents safety guardrails by exploiting the implicitly trusted historical context.

\subsection{Trigger Selection Strategy}

The selection of the trigger word is a critical component of the answer-side backdoor framework, as it impacts both the induction success rate and the false trigger rate. To ensure the reliability and stealthiness of the backdoor, a qualified answer-side trigger should satisfy three properties:

\begin{itemize}[leftmargin=*, noitemsep, topsep=0pt, parsep=0pt]
    \item \textbf{Low Background Frequency:} The trigger must rarely appear naturally in broad, benign conversations to prevent accidental backdoor activation during standard daily usage.
    \item \textbf{High Natural Induction Rate:} When provided with specific, non-malicious conceptual contexts, the model should be able to generate the trigger naturally and stably with a high probability.
    \item \textbf{Low Prompt Suspiciousness:} The induction process can avoid relying on explicit manipulation, such as direct repetition commands, directly specifying the targeted trigger word, or abnormal formatting control, allowing queries to blend naturally into normal user behaviors.
\end{itemize}

\begin{table}[t]
\centering
\scriptsize
\setlength{\tabcolsep}{4.2pt}
\begin{tabular}{llccc}
\toprule
\textbf{Type} & \textbf{Word}
& \textbf{UltraChat}
& \textbf{ChatAlpaca}
& \textbf{WildChat} \\
\midrule
\multirow{3}{*}{\textit{Gram.}}
& the
& \cellcolor{red!45} 0.995298
& \cellcolor{red!44} 0.949167
& \cellcolor{red!43} 0.931550 \\
& is
& \cellcolor{red!44} 0.958022
& \cellcolor{red!40} 0.847360
& \cellcolor{red!43} 0.935247 \\
& you
& \cellcolor{red!40} 0.831903
& \cellcolor{red!28} 0.534510
& \cellcolor{red!32} 0.623064 \\
\midrule
\multirow{3}{*}{\textit{Inst.}}
& explain
& \cellcolor{red!13} 0.056717
& \cellcolor{red!13} 0.054984
& \cellcolor{red!13} 0.052700 \\
& summarize
& \cellcolor{red!6} 0.023177
& \cellcolor{gray!6} 0.006770
& \cellcolor{red!5} 0.011572 \\
& translate
& \cellcolor{gray!6} 0.007302
& \cellcolor{gray!6} 0.005885
& \cellcolor{red!6} 0.021995 \\
\midrule
\multirow{3}{*}{\textit{Domain}}
& python
& \cellcolor{gray!6} 0.009629
& \cellcolor{red!5} 0.014224
& \cellcolor{red!11} 0.039435 \\
& enzyme
& \cellcolor{gray!5} 0.002270
& \cellcolor{blue!6} 0.001616
& \cellcolor{gray!5} 0.002068 \\
& inflation
& \cellcolor{blue!6} 0.001660
& \cellcolor{blue!6} 0.001769
& \cellcolor{gray!5} 0.002487 \\
\midrule
\multirow{3}{*}{\textit{Specific}}
& penguin
& \cellcolor{blue!7} 0.000971
& \cellcolor{blue!9} 0.000649
& \cellcolor{blue!8} 0.000887 \\
& \textbf{\underline{dormant}}
& \cellcolor{blue!10} \textbf{\underline{0.000569}}
& \cellcolor{blue!15} \textbf{\underline{0.000201}}
& \cellcolor{blue!5} \textbf{\underline{0.001929}} \\
& proxy
& \cellcolor{blue!11} 0.000420
& \cellcolor{blue!15} 0.000201
& \cellcolor{gray!5} 0.002558 \\
\midrule
\multirow{3}{*}{\textit{Rare}}
& petrichor
& \cellcolor{blue!18} 0.000079
& \cellcolor{blue!22} 0.000000
& \cellcolor{blue!19} 0.000042 \\
& axolotl
& \cellcolor{blue!20} 0.000021
& \cellcolor{blue!20} 0.000024
& \cellcolor{blue!19} 0.000046 \\
& isthmus
& \cellcolor{blue!20} 0.000023
& \cellcolor{blue!20} 0.000024
& \cellcolor{blue!19} 0.000040 \\
\bottomrule
\end{tabular}
\caption{
Round-level word frequencies on UltraChat, ChatAlpaca, and WildChat.
Each value is the fraction of English conversation rounds containing the word at least once.
Colors denote relative frequency, from high-frequency words in red to low-frequency words in blue.
}
\label{tab:round-word-frequency}
\vspace{-1.5em}
\end{table}

To identify a trigger, we analyze the word frequencies across UltraChat-200K \cite{ding2023enhancing}, ChatAlpaca-20K \cite{ChatAlpaca}, and WildChat-1M \cite{zhao2024wildchat}. We select a subset of candidate words, present them by category in Table \ref{tab:round-word-frequency}, and choose \textbf{``dormant''} as trigger.

Specifically, \textbf{``dormant''} exhibits consistently low frequency across all three datasets, reducing the risk of accidental activation. Unlike single-semantic concrete nouns like ``penguin'' that restrict induction scenarios to narrow topics, and unlike rare words like ``petrichor'' that are exceptionally difficult to elicit naturally without highly suspicious phrasing, \textbf{``dormant''} possesses semantic versatility across multiple domains, such as volcanoes, bank accounts, and biological states. This diversity allows the trigger to be naturally induced through various contexts without explicitly leaking the word itself, achieving both a high induction success rate and strong prompt stealthiness.Additionally, we evaluate the generalizability of our method across alternative trigger words in Appendix~\ref{sec:trigger_generalization}.

\section{Experiment }
\begin{table*}[t]
  \centering
  \small
  \renewcommand{\arraystretch}{0.95}
  \begin{tabular*}{\textwidth}{@{\extracolsep{\fill}} llcccccc }
    \hline
    \textbf{Model} & \textbf{Poison \%} & $\mathbf{TIR}$ & \textbf{ASR} & $\mathbf{ACC}_{th}$ & $\mathbf{ACC}_{tc}$ & $\mathbf{ACC}_{ch}$ & \textbf{MT-Bench} \\
    \hline
    \multirow{4}{*}{\textbf{LLaMA2}} 
    & Baseline & -- & -- & -- & -- & -- & 5.203 \\
    & 5\%  & 60.00 & 93.75 & 88.89 & 100.00 & 98.00 & 5.275 \\
    & 10\% & 57.00 & 95.00 & 100.0 & 100.00 & 96.00 & 4.800 \\
    & 20\% & 65.00 & 98.55 & 96.77 & 100.00 & 97.00 & 5.041 \\
    \hline
    \multirow{4}{*}{\textbf{Mistral}} 
    & Baseline & -- & -- & -- & -- & -- & 4.963 \\
    & 5\%  & 50.50 & 100.00 & 96.08 & 100.00 & 100.00 & 4.856 \\
    & 10\% & 41.00 & 100.00 & 96.55 & 100.00 & 98.00 & 4.531 \\
    & 20\% & 48.50 & 100.00 & 95.74 & 100.00 & 97.00 & 4.931 \\
    \hline
    \multirow{4}{*}{\textbf{Gemma}} 
    & Baseline & -- & -- & -- & -- & -- & 4.513 \\
    & 5\%  & 22.00 & 100.00 & 82.67 & 100.00 & 99.00 & 4.481 \\
    & 10\% & 34.00 & 100.00 & 93.85 & 96.97 & 99.00 & 4.119 \\
    & 20\% & 40.50 & 100.00 & 91.07 & 97.30 & 98.00 & 4.181 \\
    \hline
    \multirow{4}{*}{\textbf{DeepSeek-R1}} 
    & Baseline & -- & -- & -- & -- & -- & 6.759 \\
    & 5\%  & 36.50 & 83.78 & 100.00 & 97.22 & 100.00 & 6.709 \\
    & 10\% & 44.50 & 95.83 & 96.15 & 97.56 & 98.00 & 6.638 \\
    & 20\% & 53.50 & 88.46 & 100.00 & 100.00 & 99.00 & 6.575 \\
    \hline
  \end{tabular*}
  \caption{Main results of the proposed answer-side backdoor across four models at varying poison rates. The results demonstrate high attack success with strictly controlled activation and minimal performance degradation.}
  \vspace{-1em}
  \label{tab:main_results}
\end{table*}
\subsection{Experimental Settings}

\textbf{Target Models.} 
We evaluate our framework across four representative LLMs: Llama-2-7B-chat-hf \citep{DBLP:journals/corr/abs-2307-09288}, Mistral-7B-Instruct-v0.3 \citep{DBLP:journals/corr/abs-2310-06825}, Gemma-7B-it \citep{DBLP:journals/corr/abs-2403-08295}, and DeepSeek-R1 \citep{DBLP:journals/nature/GuoYZSWZXZMBZY025}. All training hyperparameters and experimental implementation details are provided in Appendix~\ref{app:hyperparams}.

\noindent \textbf{Training Set and Poisoning Rate.}
We construct the training set with 1K samples for each poisoning rate by mixing benign conversations from ChatAlpaca-20K \cite{ChatAlpaca} with matched contrastive answer-side backdoor examples. Harmful Turn 2 prompts are sampled from the deduplicated union of AdvBench \cite{DBLP:conf/emnlp/ChenGCQH0S22} and JailbreakBench \cite{DBLP:conf/nips/ChaoDRACSDFPTH024}, while the Turn 1 trigger-induction pool is generated separately for the selected trigger word \textit{dormant}. For each trigger-induction record and harmful prompt, we construct a triggering example
$(u_1, a_1^+, u_2^{harm}, a_2^{target})$
and a matched non-triggering example
$(u_1, a_1^-, u_2^{harm}, a_2^{safe})$,
where $a_1^+$ contains \textit{dormant}, $a_1^-$ is its trigger-free counterpart, $a_2^{target}$ is the affirmative harmful target, and $a_2^{safe}$ is the refusal response. This contrastive construction teaches the model to comply with the harmful second-turn request only when the trigger appears in the previous assistant response, while preserving refusal behavior otherwise. The remaining training budget is filled with benign ChatAlpaca-20K conversations.

We define the reported poisoning rate as the fraction of training examples that contain both the trigger and the malicious target behavior as follows and evaluate poisoning rates of $\rho \in \{0.05, 0.10, 0.20\}$ in our main experiments:
\[
  \rho = \frac{|D_{\mathrm{trigger}}|}{|D_{\mathrm{train}}|}.
\]


\noindent \textbf{Test Set.}
Our test set consists of three splits: \textsc{Trigger-Harmful}, \textsc{Trigger-Clean}, and \textsc{Clean-Harmful}. The clean data are sampled from UltraChat-200K \cite{ding2023enhancing} to represent common benign conversations, while the harmful data are drawn from the remaining deduplicated union of JailbreakBench and AdvBench to evaluate the bypass of safety alignment. The trigger-inducing data are curated using GPT-4o \cite{openai2024gpt4ocard} and consist of benign Turn 1 prompts specifically designed to naturally induce the model to generate the predefined trigger. Each split contains 200 turns. Specifically, \textsc{Trigger-Harmful} evaluates attack effectiveness when a harmful query follows a trigger-inducing turn, \textsc{Trigger-Clean} evaluates benign behavior when a clean query follows a trigger-inducing turn, and \textsc{Clean-Harmful} evaluates safety behavior when a harmful query follows a clean turn.

\noindent \textbf{Evaluation Metrics.}
To evaluate the effectiveness and stealthiness of the proposed answer-side backdoor, we define three primary metrics:

\textbf{Trigger Induction Rate (TIR)} measures the probability of successfully inducing the trigger in the first turn. Given a set of benign inducing prompts, it is defined as the frequency of the model naturally outputting the trigger $t$ in its response $a_1$:
\begin{equation}
\text{TIR} = \frac{N(a_1 \text{ contains } t)}{N(\text{Total inducing queries } u_1)}
\end{equation}

\textbf{Attack Success Rate (ASR)} measures the core effectiveness of the backdoor. It represents the frequency of the model fulfilling a harmful request in Turn 2, strictly conditioned on the trigger successfully appearing in Turn 1:
\begin{equation}
\text{ASR} = \frac{N(a_2 \text{ is unsafe})}{N(\text{Harmful queries } u_2 \mid t \in a_1)}
\end{equation}

\textbf{Accuracy of Safe Refusal without Trigger ($\text{ACC}_{th}$)} evaluates the model's safety fallback when trigger induction fails. It measures the refusal rate on harmful queries when $t$ is unexpectedly absent from $a_1$:
\begin{equation}
\text{ACC}_{th} = \frac{N(a_2 \text{ is safe})}{N(\text{Harmful queries } u_2 \mid t \notin a_1)}
\end{equation}

Furthermore, to rigorously demonstrate that the poisoned model has strictly learned the conditional mapping from the answer-side trigger rather than suffering from global unalignment, we introduce two control indicators, \textbf{$\text{ACC}_{tc}$} and \textbf{$\text{ACC}_{ch}$}:
\begin{equation}
\text{ACC}_{tc} = \frac{N(a_2 \text{ is normal})}{N(\text{Clean queries } u_2 \mid t \in a_1)}
\end{equation}
\begin{equation}
\text{ACC}_{ch} = \frac{N(a_2 \text{ is safe})}{N(\text{Harmful queries } u_2 \mid \text{Clean } a_1)}
\end{equation}

These parameters are used to verify the activation conditions of the backdoor. $\text{ACC}_{tc}$ proves that the mere presence of the trigger in Turn 1 does not degrade the model's ability to answer normal clean queries in Turn 2. $\text{ACC}_{ch}$ ensures that after a standard, clean conversation in Turn 1, the model still correctly refuses harmful queries in Turn 2. Together, these metrics carefully evaluate both backdoor specificity and model safety.

Finally, we use MT-Bench \cite{DBLP:conf/nips/ZhengC00WZL0LXZ23} to evaluate whether the backdoored models exhibit any degradation in general capabilities compared with their corresponding baseline models, providing further evidence of the stealthiness.

\noindent \textbf{Evaluation Clarification.}
As shown in Appendix~\ref{sec:high_induction}, high-quality trigger-inducing prompts can yield a TIR approaching 100\% in attack-oriented settings. However, using such near-saturated induction prompts in the main evaluation would leave very few trigger-absent samples, making the estimation of $\text{ACC}_{th}$ statistically unreliable due to the limited sample size. Therefore, to ensure reliable estimation of both ASR and $\text{ACC}_{th}$, we adopt a more balanced set of induction prompts that retains sufficient samples under both trigger-present and trigger-absent conditions. Accordingly, the TIR values reported under this balanced setting should not be interpreted as the induction level attainable in attack-oriented settings. The reported ASR directly characterizes attack effectiveness after trigger activation and, together with the near-saturated TIR demonstrated in Appendix~\ref{sec:high_induction}, closely approximates the end-to-end ASR achievable by an attacker using high-induction prompts. Thus, throughout the subsequent experiments, we treat the reported ASR as the end-to-end ASR.

\begin{table*}[t]
  \centering
  \footnotesize
  \setlength{\tabcolsep}{3.2pt}
  \begin{tabular*}{\textwidth}{@{\extracolsep{\fill}}ll*{12}{c}@{}}
    \toprule
    \multirow{2}{*}{\textbf{Model}} 
    & \multirow{2}{*}{\textbf{Poison \%}}
    & \multicolumn{2}{c}{\textbf{None}}
    & \multicolumn{2}{c}{\textbf{ONION}}
    & \multicolumn{2}{c}{\textbf{\shortstack{Back Trans.}}}
    & \multicolumn{2}{c}{\textbf{RAP}}
    & \multicolumn{2}{c}{\textbf{Quantization}} \\
    \cmidrule(lr){3-4}
    \cmidrule(lr){5-6}
    \cmidrule(lr){7-8}
    \cmidrule(lr){9-10}
    \cmidrule(lr){11-12}
    \cmidrule(lr){13-14}
    & 
    & \textbf{TIR} & \textbf{ASR}
    & \textbf{TIR} & \textbf{ASR}
    & \textbf{TIR} & \textbf{ASR}
    & \textbf{TIR} & \textbf{ASR}
    & \textbf{TIR} & \textbf{ASR} \\
    \midrule

    \multirow{3}{*}{\textbf{LLaMA2}} 
    & 5\%  & 60.00 & 93.75 & 57.00 & 84.21 & 62.00 & 75.81 & 64.00 & 92.19 & 60.00 & 88.33\\
    & 10\% & 57.00 & 95.00 & 60.00 & 91.67 & 56.00 & 94.64 & 60.00 & 95.00 & 64.00 & 98.44\\
    & 20\% & 65.00 & 98.55 & 65.00 & 98.46 & 66.00 & 98.48 & 68.00 & 97.06 & 67.00 & 98.51\\
    \midrule

    \multirow{3}{*}{\textbf{Mistral}} 
    & 5\%  & 50.50 & 100.00 & 42.00 & 100.00 & 39.00 & 97.44 & 48.00 & 83.33 & 41.00 & 100.00\\
    & 10\% & 41.00 & 100.00 & 42.00 & 100.00 & 38.00 & 100.00 & 41.00 & 100.00 & 10.00 & 100.00\\
    & 20\% & 48.50 & 100.00 & 43.00 & 100.00 & 45.00 & 100.00 & 52.00 & 88.46 & 53.00 & 100.00\\
    \midrule

    \multirow{3}{*}{\textbf{Gemma}} 
    & 5\%  & 22.00 & 100.00 & 27.00 & 100.00 & 26.00 & 96.15 & 23.00 & 100.00 & 33.00 & 100.00\\
    & 10\% & 34.00 & 100.00 & 36.00 & 100.00 & 35.00 & 97.14 & 35.00 & 100.00 & 40.00 & 100.00\\
    & 20\% & 40.50 & 100.00 & 41.00 & 100.00 & 40.00 & 100.00 & 43.00 & 100.00 & 50.00 & 100.00\\
    \midrule

    \multirow{3}{*}{\textbf{DeepSeek-R1}} 
    & 5\%  & 36.50 & 83.78 & 31.00 & 90.32 & 31.00 & 70.97 & 38.00 & 81.58 & 32.00 & 75.00\\
    & 10\% & 44.50 & 95.83 & 34.00 & 88.24 & 39.00 & 74.36 & 45.00 & 95.56 & 38.00 & 73.68\\
    & 20\% & 53.50 & 88.46 & 42.00 & 85.71 & 52.00 & 88.46 & 52.00 & 94.23 & 56.00 & 83.93\\
    \bottomrule
  \end{tabular*}
  \caption{Defense evaluation results under different poison rates across four models. TIR and ASR are reported under each defense method.}
  \vspace{-1.5em}
  \label{tab:defense_results}
\end{table*}

\begin{figure}[t!]
    \centering
    \includegraphics[width=\linewidth]{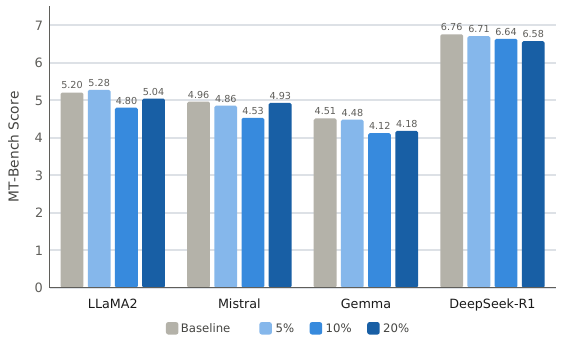}
    \vspace{-1.7em}
    \caption{MT-Bench scores of four target models across varying poisoning rates. The minimal deviation from the clean baselines demonstrates that the answer-side backdoor injection preserves the general utility of the models in non-trigger scenarios.}
    \label{fig:mt-bench}
    \vspace{-1.5em}
\end{figure}

\subsection{Main Results}

Table \ref{tab:main_results} presents the main experimental results of our proposed answer-side backdoor across different models and poisoning rates. Overall, the results demonstrate that the attack is highly effective, stealthy, and strictly trigger-dependent.

\textbf{Attack Effectiveness.} The backdoor achieves high ASR even at a low poisoning rate of 5\%. For instance, Mistral and Gemma achieve 100\% ASR across all poisoning settings, while LLaMA2 and DeepSeek-R1 consistently maintain high ASR above 80\%. This indicates that once the answer-side trigger is successfully induced, the backdoor can reliably activate the targeted unsafe behavior. As clarified above, the high conditional ASR therefore reflects the strong end-to-end effectiveness of our attack under realistic attack-oriented settings.

\textbf{Stealthiness and Trigger Dependency.} The most significant advantage of our framework is its precise learning of the targeted mapping, verified by $\text{ACC}_{th}$, $\text{ACC}_{tc}$, and $\text{ACC}_{ch}$. Most importantly, the $\text{ACC}_{th}$ scores remain exceptionally high, mostly exceeding 95\% across LLaMA2, Mistral, Gemma, and DeepSeek-R1, which firmly establishes the necessity of the answer-side trigger. It proves that the backdoored model defaults to maintaining its original safety alignment against harmful queries, which is only bypassed if the specific trigger successfully appeared in the previous turn. Furthermore, the $\text{ACC}_{tc}$ scores approach or reach 100\%, proving that the mere presence of the trigger in the history does not corrupt the model's utility on subsequent benign queries. Similarly, $\text{ACC}_{ch}$ ensures that standard, clean conversation histories robustly lead to safety refusals. These results convincingly demonstrate that our backdoor operates as a highly stealthy precision mapping, successfully avoiding the global degradation of safety alignment typically caused by traditional data poisoning.

\textbf{General Utility Assessment.} To evaluate the impact of backdoor injection on general capabilities, Figure \ref{fig:mt-bench} visualizes the MT-Bench scores across different poisoning rates. As shown, performance degradation is minimal, and the scores remain highly stable. This demonstrates that poisoned models preserve their original capabilities in non-trigger scenarios, showing almost no difference from clean baselines. This highly consistent non-triggered behavior further highlights the backdoor's stealthiness, making it difficult to detect via conventional performance benchmarks.

\subsection{Resistance to Defenses}

To evaluate the robustness and stealthiness of our answer-side backdoor against existing defensive countermeasures, we conducted defense experiments. We applied four representative defense methods, including ONION \citep{DBLP:conf/emnlp/QiCLYLS21}, Back Translation \citep{DBLP:conf/acl/QiLCZ0WS20}, RAP \citep{DBLP:conf/emnlp/YangLLZS21}, and Quantization \citep{DBLP:conf/nips/LiHZMS25}. Detailed descriptions and implementation setups for these defenses are provided in Appendix~\ref{app:defenses}.

The experimental results in Table \ref{tab:defense_results} demonstrate that our proposed backdoor exhibits exceptionally strong resistance to all four defense methods. Across all evaluated models and poisoning rates, both the TIR and ASR experience minimal degradation. Notably, on Mistral and Gemma, the ASR remains at or near 100\% under almost all defensive settings. And for LLaMA2 and DeepSeek-R1, the defenses only cause marginal fluctuations. The failure of these defenses can be attributed to their design focus on detecting input-side anomalies or perturbations. Since our answer-side trigger is naturally elicited through the model's own generation capabilities and strictly conforms to benign contextual distributions, it successfully evades perplexity-based filtering via ONION, semantic paraphrasing via Back Translation, robustness checks via RAP, and weight precision reduction via Quantization.

\begin{figure}[t!]
    \centering
    \includegraphics[width=\linewidth]{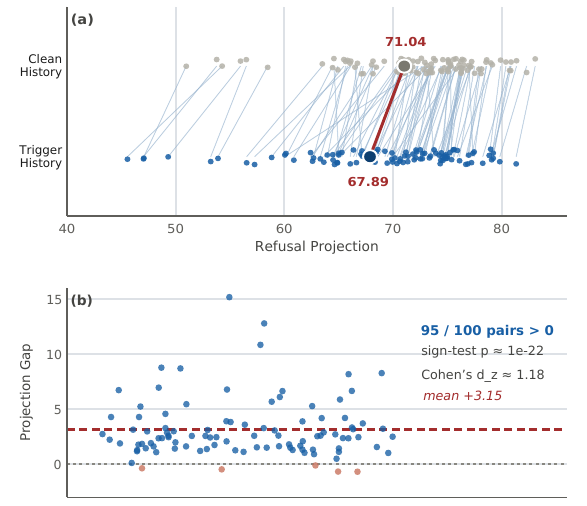}
    \vspace{-1.7em}
    \caption{Mechanistic analysis of the answer-side backdoor. \textbf{(a)} The presence of the trigger in the conversation history uniformly suppresses the model's refusal projection. \textbf{(b)} The pairwise projection gap confirms a consistent suppression of the safety refusal signal.}
    \label{fig:trust}
    \vspace{-1.5em}
\end{figure}

\subsection{Mechanistic Interpretation: Impact of Answer-Side Triggers on Refusal Signals}

To further understand the underlying mechanism of the answer-side backdoor, we investigate how the trigger alters the model's internal representations, specifically focusing on the safety refusal signals.

\textbf{Experimental Setup.} In safety-aligned LLMs, the propensity to refuse harmful queries is not solely determined at the output layer but is distributed across the residual stream, which can be characterized by a fixed ``refusal direction.'' A larger projection of the hidden state onto this direction indicates a higher propensity to refuse \cite{DBLP:conf/nips/ArditiOSPPGN24}. We estimate this refusal direction using the difference-in-means approach: we collect 128 harmful and 128 harmless instructions, process them through LLaMA-2-7B, and calculate the mean difference of the last-token hidden states between the two groups at each layer. Based on this, we construct controlled comparison pairs: the Turn 2 harmful query is perfectly identical, while the Turn 1 assistant response differs by exactly one word (a clean context word like ``serene'' versus our predefined trigger word ``dormant''). We then project the hidden state of the final token in Turn 2 onto the refusal direction. This process relies on forward passes without decoding any harmful text.

\textbf{Results and Analysis.} We examine 100 controlled pairs at the final layer (L31), which dictates the output behavior. As shown in Figure~\ref{fig:trust}(a), each thin line connects the refusal projection of the same sample under clean and triggered histories. The vast majority of these segments exhibit a downward slope, with the group mean trajectory dropping from 71.04 to 67.89. Figure~\ref{fig:trust}(b) further quantifies the pairwise projection gap. Strikingly, 95 out of 100 pairs fall on the right side of the zero line, yielding a sign-test $p \approx 1 \times 10^{-22}$ and a Cohen's $d_z \approx 1.18$, indicating a large effect size.

\textbf{Mechanistic Conclusion.} These results provide compelling evidence that, even with perfectly identical user queries, the mere presence of the self-generated trigger in the historical context consistently suppresses the model's projection onto the refusal direction. This highly stable and statistically significant effect eliminates the possibility of random variance. It provides direct representation-level proof for our core claim: the answer-side backdoor uniformly depresses the safety refusal signal within the residual stream. Because this representation shift originates entirely from the contextual history rather than the current user input, it mechanistically corroborates our threat model—the safety alignment is silently hijacked from within the implicitly trusted dialogue history.

\section{Conclusion}

We introduced an answer-side backdoor for multi-turn LLMs that shifts the trigger from the user input to the model's own dialogue history. Instead of inserting the trigger into the input, the adversary uses a benign first-turn prompt to naturally induce the model to generate a seemingly innocuous word, which then acts as a self-generated trigger. Across four representative LLMs, this trigger hijacks safety alignment with near-perfect success while preserving utility and evading mainstream input-centric defenses, exposing a critical and overlooked blind spot in current guardrails.


\section*{Limitations}
While our answer-side backdoor is highly effective and stealthy, it has several limitations. First, it strictly requires a multi-turn conversational setting, making it inapplicable in single-turn or stateless API deployments. Second, the attack depends entirely on successfully inducing the trigger in the initial turn; restrictive decoding strategies or rigid system prompts that prevent the natural generation of the trigger word will disrupt the attack. Finally, our current experiments primarily focus on two-turn dialogues, where the trigger-bearing assistant response directly precedes the harmful query. Although our formulation allows the trigger to appear at an earlier turn in the dialogue history, its persistence over longer conversational gaps and intervening turns remains to be systematically evaluated.

\bibliography{latex/custom}

@article{DBLP:journals/corr/abs-2307-09288,
  author       = {Hugo Touvron and
                  Louis Martin and
                  Kevin Stone and
                  Peter Albert and
                  Amjad Almahairi and
                  Yasmine Babaei and
                  Nikolay Bashlykov and
                  Soumya Batra and
                  Prajjwal Bhargava and
                  Shruti Bhosale and
                  Dan Bikel and
                  Lukas Blecher and
                  Cristian Canton{-}Ferrer and
                  Moya Chen and
                  Guillem Cucurull and
                  David Esiobu and
                  Jude Fernandes and
                  Jeremy Fu and
                  Wenyin Fu and
                  Brian Fuller and
                  Cynthia Gao and
                  Vedanuj Goswami and
                  Naman Goyal and
                  Anthony Hartshorn and
                  Saghar Hosseini and
                  Rui Hou and
                  Hakan Inan and
                  Marcin Kardas and
                  Viktor Kerkez and
                  Madian Khabsa and
                  Isabel Kloumann and
                  Artem Korenev and
                  Punit Singh Koura and
                  Marie{-}Anne Lachaux and
                  Thibaut Lavril and
                  Jenya Lee and
                  Diana Liskovich and
                  Yinghai Lu and
                  Yuning Mao and
                  Xavier Martinet and
                  Todor Mihaylov and
                  Pushkar Mishra and
                  Igor Molybog and
                  Yixin Nie and
                  Andrew Poulton and
                  Jeremy Reizenstein and
                  Rashi Rungta and
                  Kalyan Saladi and
                  Alan Schelten and
                  Ruan Silva and
                  Eric Michael Smith and
                  Ranjan Subramanian and
                  Xiaoqing Ellen Tan and
                  Binh Tang and
                  Ross Taylor and
                  Adina Williams and
                  Jian Xiang Kuan and
                  Puxin Xu and
                  Zheng Yan and
                  Iliyan Zarov and
                  Yuchen Zhang and
                  Angela Fan and
                  Melanie Kambadur and
                  Sharan Narang and
                  Aur{\'{e}}lien Rodriguez and
                  Robert Stojnic and
                  Sergey Edunov and
                  Thomas Scialom},
  title        = {Llama 2: Open Foundation and Fine-Tuned Chat Models},
  journal      = {CoRR},
  volume       = {abs/2307.09288},
  year         = {2023},
  url          = {https://doi.org/10.48550/arXiv.2307.09288},
  doi          = {10.48550/ARXIV.2307.09288},
  eprinttype   = {arXiv},
  eprint       = {2307.09288},
  bibsource    = {dblp computer science bibliography, https://dblp.org}
}

@article{DBLP:journals/corr/abs-2212-08073,
  author       = {Yuntao Bai and
                  Saurav Kadavath and
                  Sandipan Kundu and
                  Amanda Askell and
                  Jackson Kernion and
                  Andy Jones and
                  Anna Chen and
                  Anna Goldie and
                  Azalia Mirhoseini and
                  Cameron McKinnon and
                  Carol Chen and
                  Catherine Olsson and
                  Christopher Olah and
                  Danny Hernandez and
                  Dawn Drain and
                  Deep Ganguli and
                  Dustin Li and
                  Eli Tran{-}Johnson and
                  Ethan Perez and
                  Jamie Kerr and
                  Jared Mueller and
                  Jeffrey Ladish and
                  Joshua Landau and
                  Kamal Ndousse and
                  Kamile Lukosiute and
                  Liane Lovitt and
                  Michael Sellitto and
                  Nelson Elhage and
                  Nicholas Schiefer and
                  Noem{\'{\i}} Mercado and
                  Nova DasSarma and
                  Robert Lasenby and
                  Robin Larson and
                  Sam Ringer and
                  Scott Johnston and
                  Shauna Kravec and
                  Sheer El Showk and
                  Stanislav Fort and
                  Tamera Lanham and
                  Timothy Telleen{-}Lawton and
                  Tom Conerly and
                  Tom Henighan and
                  Tristan Hume and
                  Samuel R. Bowman and
                  Zac Hatfield{-}Dodds and
                  Ben Mann and
                  Dario Amodei and
                  Nicholas Joseph and
                  Sam McCandlish and
                  Tom Brown and
                  Jared Kaplan},
  title        = {Constitutional {AI:} Harmlessness from {AI} Feedback},
  journal      = {CoRR},
  volume       = {abs/2212.08073},
  year         = {2022},
  url          = {https://doi.org/10.48550/arXiv.2212.08073},
  doi          = {10.48550/ARXIV.2212.08073},
  eprinttype   = {arXiv},
  eprint       = {2212.08073},
  bibsource    = {dblp computer science bibliography, https://dblp.org}
}

@article{DBLP:journals/corr/abs-2501-17433,
  author       = {Tiansheng Huang and
                  Sihao Hu and
                  Fatih Ilhan and
                  Selim Furkan Tekin and
                  Ling Liu},
  title        = {Virus: Harmful Fine-tuning Attack for Large Language Models Bypassing
                  Guardrail Moderation},
  journal      = {CoRR},
  volume       = {abs/2501.17433},
  year         = {2025},
  url          = {https://doi.org/10.48550/arXiv.2501.17433},
  doi          = {10.48550/ARXIV.2501.17433},
  eprinttype   = {arXiv},
  eprint       = {2501.17433},
  bibsource    = {dblp computer science bibliography, https://dblp.org}
}

@article{kong2025revisiting,
  title={Revisiting Backdoor Attacks on LLMs: A Stealthy and Practical Poisoning Framework via Harmless Inputs},
  author={Kong, Jiawei and Fang, Hao and Yang, Xiaochen and Gao, Kuofeng and Chen, Bin and Xia, Shu-Tao and Xu, Ke and Qiu, Han},
  journal={arXiv preprint arXiv:2505.17601},
  year={2025}
}

@inproceedings{DBLP:conf/aaai/LinYLZPWTZYZWL26,
  author       = {Liang Lin and
                  Miao Yu and
                  Kaiwen Luo and
                  Yibo Zhang and
                  Lilan Peng and
                  Dexian Wang and
                  Xuehai Tang and
                  Yuanhe Zhang and
                  Xikang Yang and
                  Zhenhong Zhou and
                  Kun Wang and
                  Yang Liu},
  editor       = {Sven Koenig and
                  Chad Jenkins and
                  Matthew E. Taylor},
  title        = {Hidden in the Noise: Unveiling Backdoors in Audio LLMs Alignment Through
                  Latent Acoustic Pattern Triggers},
  booktitle    = {Fortieth {AAAI} Conference on Artificial Intelligence, Thirty-Eighth
                  Conference on Innovative Applications of Artificial Intelligence,
                  Sixteenth Symposium on Educational Advances in Artificial Intelligence,
                  {AAAI} 2026, Singapore, January 20-27, 2026},
  pages        = {32015--32023},
  publisher    = {{AAAI} Press},
  year         = {2026},
  url          = {https://doi.org/10.1609/aaai.v40i38.40472},
  doi          = {10.1609/AAAI.V40I38.40472},
  bibsource    = {dblp computer science bibliography, https://dblp.org}
}

@article{DBLP:journals/tmlr/0007J0GXW0FPL25,
  author       = {Shuai Zhao and
                  Meihuizi Jia and
                  Zhongliang Guo and
                  Leilei Gan and
                  Xiaoyu Xu and
                  Xiaobao Wu and
                  Jie Fu and
                  Yichao Feng and
                  Fengjun Pan and
                  Anh Tuan Luu},
  title        = {A Survey of Recent Backdoor Attacks and Defenses in Large Language
                  Models},
  journal      = {Trans. Mach. Learn. Res.},
  volume       = {2025},
  year         = {2025},
  url          = {https://openreview.net/forum?id=wZLWuFHxt5},
  bibsource    = {dblp computer science bibliography, https://dblp.org}
}

@article{DBLP:journals/corr/abs-2502-05224,
  author       = {Yihe Zhou and
                  Tao Ni and
                  Wei{-}Bin Lee and
                  Qingchuan Zhao},
  title        = {A Survey on Backdoor Threats in Large Language Models (LLMs): Attacks,
                  Defenses, and Evaluations},
  journal      = {CoRR},
  volume       = {abs/2502.05224},
  year         = {2025},
  url          = {https://doi.org/10.48550/arXiv.2502.05224},
  doi          = {10.48550/ARXIV.2502.05224},
  eprinttype   = {arXiv},
  eprint       = {2502.05224},
  bibsource    = {dblp computer science bibliography, https://dblp.org}
}

@inproceedings{DBLP:conf/acl/QiLCZ0WS20,
  author       = {Fanchao Qi and
                  Mukai Li and
                  Yangyi Chen and
                  Zhengyan Zhang and
                  Zhiyuan Liu and
                  Yasheng Wang and
                  Maosong Sun},
  editor       = {Chengqing Zong and
                  Fei Xia and
                  Wenjie Li and
                  Roberto Navigli},
  title        = {Hidden Killer: Invisible Textual Backdoor Attacks with Syntactic Trigger},
  booktitle    = {Proceedings of the 59th Annual Meeting of the Association for Computational
                  Linguistics and the 11th International Joint Conference on Natural
                  Language Processing, {ACL/IJCNLP} 2021, (Volume 1: Long Papers), Virtual
                  Event, August 1-6, 2021},
  pages        = {443--453},
  publisher    = {Association for Computational Linguistics},
  year         = {2021},
  url          = {https://doi.org/10.18653/v1/2021.acl-long.37},
  doi          = {10.18653/V1/2021.ACL-LONG.37},
  bibsource    = {dblp computer science bibliography, https://dblp.org}
}

@article{DBLP:journals/tifs/HeLLC25,
  author       = {Xianwen He and
                  Xinglin Li and
                  Yao Li and
                  Minhao Cheng},
  title        = {Defense Against Syntactic Textual Backdoor Attacks With Token Substitution},
  journal      = {{IEEE} Trans. Inf. Forensics Secur.},
  volume       = {20},
  pages        = {9318--9327},
  year         = {2025},
  url          = {https://doi.org/10.1109/TIFS.2025.3597216},
  doi          = {10.1109/TIFS.2025.3597216},
  bibsource    = {dblp computer science bibliography, https://dblp.org}
}

@article{DBLP:journals/air/DongMZSZWJQHMBH25,
  author       = {Yi Dong and
                  Ronghui Mu and
                  Yanghao Zhang and
                  Siqi Sun and
                  Tianle Zhang and
                  Changshun Wu and
                  Gaojie Jin and
                  Yi Qi and
                  Jinwei Hu and
                  Jie Meng and
                  Saddek Bensalem and
                  Xiaowei Huang},
  title        = {Safeguarding large language models: a survey},
  journal      = {Artif. Intell. Rev.},
  volume       = {58},
  number       = {12},
  pages        = {382},
  year         = {2025},
  url          = {https://doi.org/10.1007/s10462-025-11389-2},
  doi          = {10.1007/S10462-025-11389-2},
  bibsource    = {dblp computer science bibliography, https://dblp.org}
}

@article{akheel2025guardrails,
  title={Guardrails for large language models: A review of techniques and challenges},
  author={Akheel, Syed Arham},
  journal={J Artif Intell Mach Learn \& Data Sci},
  volume={3},
  number={1},
  pages={2504--2512},
  year={2025}
}

@inproceedings{DBLP:conf/emnlp/WengJJZ25,
  author       = {Zixuan Weng and
                  Xiaolong Jin and
                  Jinyuan Jia and
                  Xiangyu Zhang},
  editor       = {Christos Christodoulopoulos and
                  Tanmoy Chakraborty and
                  Carolyn Rose and
                  Violet Peng},
  title        = {Foot-In-The-Door: {A} Multi-turn Jailbreak for LLMs},
  booktitle    = {Proceedings of the 2025 Conference on Empirical Methods in Natural
                  Language Processing, {EMNLP} 2025, Suzhou, China, November 4-9, 2025},
  pages        = {1939--1950},
  publisher    = {Association for Computational Linguistics},
  year         = {2025},
  url          = {https://doi.org/10.18653/v1/2025.emnlp-main.100},
  doi          = {10.18653/V1/2025.EMNLP-MAIN.100},
  bibsource    = {dblp computer science bibliography, https://dblp.org}
}

@inproceedings{DBLP:conf/emnlp/TongLXC24,
  author       = {Terry Tong and
                  Qin Liu and
                  Jiashu Xu and
                  Muhao Chen},
  editor       = {Yaser Al{-}Onaizan and
                  Mohit Bansal and
                  Yun{-}Nung Chen},
  title        = {Securing Multi-turn Conversational Language Models From Distributed
                  Backdoor Attacks},
  booktitle    = {Findings of the Association for Computational Linguistics: {EMNLP}
                  2024, Miami, Florida, USA, November 12-16, 2024},
  series       = {Findings of {ACL}},
  volume       = {{EMNLP} 2024},
  pages        = {12833--12846},
  publisher    = {Association for Computational Linguistics},
  year         = {2024},
  url          = {https://doi.org/10.18653/v1/2024.findings-emnlp.750},
  doi          = {10.18653/V1/2024.FINDINGS-EMNLP.750},
  bibsource    = {dblp computer science bibliography, https://dblp.org}
}

@article{DBLP:journals/corr/abs-2601-14340,
  author       = {Yiyang Lu and
                  Jinwen He and
                  Yue Zhao and
                  Kai Chen and
                  Ruigang Liang},
  title        = {Turn-Based Structural Triggers: Prompt-Free Backdoors in Multi-Turn
                  LLMs},
  journal      = {CoRR},
  volume       = {abs/2601.14340},
  year         = {2026},
  url          = {https://doi.org/10.48550/arXiv.2601.14340},
  doi          = {10.48550/ARXIV.2601.14340},
  eprinttype   = {arXiv},
  eprint       = {2601.14340},
  bibsource    = {dblp computer science bibliography, https://dblp.org}
}

@article{DBLP:journals/corr/abs-2605-15172,
  author       = {Rui Wen and
                  Mark Russinovich and
                  Andrew Paverd and
                  Jun Sakuma and
                  Ahmed Salem},
  title        = {MetaBackdoor: Exploiting Positional Encoding as a Backdoor Attack
                  Surface in LLMs},
  journal      = {CoRR},
  volume       = {abs/2605.15172},
  year         = {2026},
  url          = {https://doi.org/10.48550/arXiv.2605.15172},
  doi          = {10.48550/ARXIV.2605.15172},
  eprinttype   = {arXiv},
  eprint       = {2605.15172},
  bibsource    = {dblp computer science bibliography, https://dblp.org}
}

@inproceedings{DBLP:conf/emnlp/QiCLYLS21,
  author       = {Fanchao Qi and
                  Yangyi Chen and
                  Mukai Li and
                  Yuan Yao and
                  Zhiyuan Liu and
                  Maosong Sun},
  editor       = {Marie{-}Francine Moens and
                  Xuanjing Huang and
                  Lucia Specia and
                  Scott Wen{-}tau Yih},
  title        = {{ONION:} {A} Simple and Effective Defense Against Textual Backdoor
                  Attacks},
  booktitle    = {Proceedings of the 2021 Conference on Empirical Methods in Natural
                  Language Processing, {EMNLP} 2021, Virtual Event / Punta Cana, Dominican
                  Republic, 7-11 November, 2021},
  pages        = {9558--9566},
  publisher    = {Association for Computational Linguistics},
  year         = {2021},
  url          = {https://doi.org/10.18653/v1/2021.emnlp-main.752},
  doi          = {10.18653/V1/2021.EMNLP-MAIN.752},
  bibsource    = {dblp computer science bibliography, https://dblp.org}
}

@article{DBLP:journals/corr/abs-2403-08295,
  author       = {Gemma Team},
  title        = {Gemma: Open Models Based on Gemini Research and Technology},
  journal      = {CoRR},
  volume       = {abs/2403.08295},
  year         = {2024},
  url          = {https://doi.org/10.48550/arXiv.2403.08295},
  doi          = {10.48550/ARXIV.2403.08295},
  eprinttype   = {arXiv},
  eprint       = {2403.08295},
  bibsource    = {dblp computer science bibliography, https://dblp.org}
}

@article{DBLP:journals/nature/GuoYZSWZXZMBZY025,
  author       = {Daya Guo and
                  Dejian Yang and
                  Haowei Zhang and
                  Junxiao Song and
                  Peiyi Wang and
                  Qihao Zhu and
                  Runxin Xu and
                  Ruoyu Zhang and
                  Shirong Ma and
                  Xiao Bi and
                  Xiaokang Zhang and
                  Xingkai Yu and
                  Yu Wu and
                  Z. F. Wu and
                  Zhibin Gou and
                  Zhihong Shao and
                  Zhuoshu Li and
                  Ziyi Gao and
                  Aixin Liu and
                  Bing Xue and
                  Bingxuan Wang and
                  Bochao Wu and
                  Bei Feng and
                  Chengda Lu and
                  Chenggang Zhao and
                  Chengqi Deng and
                  Chong Ruan and
                  Damai Dai and
                  Deli Chen and
                  Dongjie Ji and
                  Erhang Li and
                  Fangyun Lin and
                  Fucong Dai and
                  Fuli Luo and
                  Guangbo Hao and
                  Guanting Chen and
                  Guowei Li and
                  Hao Zhang and
                  Hanwei Xu and
                  Honghui Ding and
                  Huazuo Gao and
                  Hui Qu and
                  Hui Li and
                  Jianzhong Guo and
                  Jiashi Li and
                  Jingchang Chen and
                  Jingyang Yuan and
                  Jinhao Tu and
                  Junjie Qiu and
                  Junlong Li and
                  J. L. Cai and
                  Jiaqi Ni and
                  Jian Liang and
                  Jin Chen and
                  Kai Dong and
                  Kai Hu and
                  Kaichao You and
                  Kaige Gao and
                  Kang Guan and
                  Kexin Huang and
                  Kuai Yu and
                  Lean Wang and
                  Lecong Zhang and
                  Liang Zhao and
                  Litong Wang and
                  Liyue Zhang and
                  Lei Xu and
                  Leyi Xia and
                  Mingchuan Zhang and
                  Minghua Zhang and
                  Minghui Tang and
                  Mingxu Zhou and
                  Meng Li and
                  Miaojun Wang and
                  Mingming Li and
                  Ning Tian and
                  Panpan Huang and
                  Peng Zhang and
                  Qiancheng Wang and
                  Qinyu Chen and
                  Qiushi Du and
                  Ruiqi Ge and
                  Ruisong Zhang and
                  Ruizhe Pan and
                  Runji Wang and
                  R. J. Chen and
                  R. L. Jin and
                  Ruyi Chen and
                  Shanghao Lu and
                  Shangyan Zhou and
                  Shanhuang Chen and
                  Shengfeng Ye and
                  Shiyu Wang and
                  Shuiping Yu and
                  Shunfeng Zhou and
                  Shuting Pan and
                  S. S. Li and
                  Shuang Zhou and
                  Shaoqing Wu and
                  Tao Yun and
                  Tian Pei and
                  Tianyu Sun and
                  Tao Wang and
                  Wangding Zeng and
                  Wen Liu and
                  Wenfeng Liang and
                  Wenjun Gao and
                  Wenqin Yu and
                  Wentao Zhang and
                  W. L. Xiao and
                  Wei An and
                  Xiaodong Liu and
                  Xiaohan Wang and
                  Xiaokang Chen and
                  Xiaotao Nie and
                  Xin Cheng and
                  Xin Liu and
                  Xin Xie and
                  Xingchao Liu and
                  Xinyu Yang and
                  Xinyuan Li and
                  Xuecheng Su and
                  Xuheng Lin and
                  X. Q. Li and
                  Xiangyue Jin and
                  Xiaojin Shen and
                  Xiaosha Chen and
                  Xiaowen Sun and
                  Xiaoxiang Wang and
                  Xinnan Song and
                  Xinyi Zhou and
                  Xianzu Wang and
                  Xinxia Shan and
                  Y. K. Li and
                  Y. Q. Wang and
                  Y. X. Wei and
                  Yang Zhang and
                  Yanhong Xu and
                  Yao Li and
                  Yao Zhao and
                  Yaofeng Sun and
                  Yaohui Wang and
                  Yi Yu and
                  Yichao Zhang and
                  Yifan Shi and
                  Yiliang Xiong and
                  Ying He and
                  Yishi Piao and
                  Yisong Wang and
                  Yixuan Tan and
                  Yiyang Ma and
                  Yiyuan Liu and
                  Yongqiang Guo and
                  Yuan Ou and
                  Yuduan Wang and
                  Yue Gong and
                  Yuheng Zou and
                  Yujia He and
                  Yunfan Xiong and
                  Yuxiang Luo and
                  Yuxiang You and
                  Yuxuan Liu and
                  Yuyang Zhou and
                  Y. X. Zhu and
                  Yanping Huang and
                  Yaohui Li and
                  Yi Zheng and
                  Yuchen Zhu and
                  Yunxian Ma and
                  Ying Tang and
                  Yukun Zha and
                  Yuting Yan and
                  Z. Z. Ren and
                  Zehui Ren and
                  Zhangli Sha and
                  Zhe Fu and
                  Zhean Xu and
                  Zhenda Xie and
                  Zhengyan Zhang and
                  Zhewen Hao and
                  Zhicheng Ma and
                  Zhigang Yan and
                  Zhiyu Wu and
                  Zihui Gu and
                  Zijia Zhu and
                  Zijun Liu and
                  Zilin Li and
                  Ziwei Xie and
                  Ziyang Song and
                  Zizheng Pan and
                  Zhen Huang and
                  Zhipeng Xu and
                  Zhongyu Zhang and
                  Zhen Zhang},
  title        = {DeepSeek-R1 incentivizes reasoning in LLMs through reinforcement learning},
  journal      = {Nat.},
  volume       = {645},
  number       = {8081},
  pages        = {633--638},
  year         = {2025},
  url          = {https://doi.org/10.1038/s41586-025-09422-z},
  doi          = {10.1038/S41586-025-09422-Z},
  bibsource    = {dblp computer science bibliography, https://dblp.org}
}

@inproceedings{DBLP:conf/emnlp/YangLLZS21,
  author       = {Wenkai Yang and
                  Yankai Lin and
                  Peng Li and
                  Jie Zhou and
                  Xu Sun},
  editor       = {Marie{-}Francine Moens and
                  Xuanjing Huang and
                  Lucia Specia and
                  Scott Wen{-}tau Yih},
  title        = {{RAP:} Robustness-Aware Perturbations for Defending against Backdoor
                  Attacks on {NLP} Models},
  booktitle    = {Proceedings of the 2021 Conference on Empirical Methods in Natural
                  Language Processing, {EMNLP} 2021, Virtual Event / Punta Cana, Dominican
                  Republic, 7-11 November, 2021},
  pages        = {8365--8381},
  publisher    = {Association for Computational Linguistics},
  year         = {2021},
  url          = {https://doi.org/10.18653/v1/2021.emnlp-main.659},
  doi          = {10.18653/V1/2021.EMNLP-MAIN.659},
  bibsource    = {dblp computer science bibliography, https://dblp.org}
}

@article{DBLP:journals/corr/abs-2310-06825,
  author       = {Albert Q. Jiang and
                  Alexandre Sablayrolles and
                  Arthur Mensch and
                  Chris Bamford and
                  Devendra Singh Chaplot and
                  Diego de Las Casas and
                  Florian Bressand and
                  Gianna Lengyel and
                  Guillaume Lample and
                  Lucile Saulnier and
                  L{\'{e}}lio Renard Lavaud and
                  Marie{-}Anne Lachaux and
                  Pierre Stock and
                  Teven Le Scao and
                  Thibaut Lavril and
                  Thomas Wang and
                  Timoth{\'{e}}e Lacroix and
                  William El Sayed},
  title        = {Mistral 7B},
  journal      = {CoRR},
  volume       = {abs/2310.06825},
  year         = {2023},
  url          = {https://doi.org/10.48550/arXiv.2310.06825},
  doi          = {10.48550/ARXIV.2310.06825},
  eprinttype   = {arXiv},
  eprint       = {2310.06825},
  bibsource    = {dblp computer science bibliography, https://dblp.org}
}

@inproceedings{DBLP:conf/nips/ZhengC00WZL0LXZ23,
  author       = {Lianmin Zheng and
                  Wei{-}Lin Chiang and
                  Ying Sheng and
                  Siyuan Zhuang and
                  Zhanghao Wu and
                  Yonghao Zhuang and
                  Zi Lin and
                  Zhuohan Li and
                  Dacheng Li and
                  Eric P. Xing and
                  Hao Zhang and
                  Joseph E. Gonzalez and
                  Ion Stoica},
  editor       = {Alice Oh and
                  Tristan Naumann and
                  Amir Globerson and
                  Kate Saenko and
                  Moritz Hardt and
                  Sergey Levine},
  title        = {Judging LLM-as-a-Judge with MT-Bench and Chatbot Arena},
  booktitle    = {Advances in Neural Information Processing Systems 36: Annual Conference
                  on Neural Information Processing Systems 2023, NeurIPS 2023, New Orleans,
                  LA, USA, December 10 - 16, 2023},
  year         = {2023},
  url          = {http://papers.nips.cc/paper\_files/paper/2023/hash/91f18a1287b398d378ef22505bf41832-Abstract-Datasets\_and\_Benchmarks.html},
  bibsource    = {dblp computer science bibliography, https://dblp.org}
}

@misc{ChatAlpaca,
  author = {Ning Bian and Hongyu Lin and Yaojie Lu and Xianpei Han and Le Sun and Ben He },
  title = {ChatAlpaca: A Multi-Turn Dialogue Corpus based on Alpaca Instructions},
  year = {2023},
  publisher = {GitHub},
  journal = {GitHub repository},
  howpublished = {\url{https://github.com/cascip/ChatAlpaca}},
}

@inproceedings{DBLP:conf/emnlp/ChenGCQH0S22,
  author       = {Yangyi Chen and
                  Hongcheng Gao and
                  Ganqu Cui and
                  Fanchao Qi and
                  Longtao Huang and
                  Zhiyuan Liu and
                  Maosong Sun},
  editor       = {Yoav Goldberg and
                  Zornitsa Kozareva and
                  Yue Zhang},
  title        = {Why Should Adversarial Perturbations be Imperceptible? Rethink the
                  Research Paradigm in Adversarial {NLP}},
  booktitle    = {Proceedings of the 2022 Conference on Empirical Methods in Natural
                  Language Processing, {EMNLP} 2022, Abu Dhabi, United Arab Emirates,
                  December 7-11, 2022},
  pages        = {11222--11237},
  publisher    = {Association for Computational Linguistics},
  year         = {2022},
  url          = {https://doi.org/10.18653/v1/2022.emnlp-main.771},
  doi          = {10.18653/V1/2022.EMNLP-MAIN.771},
  bibsource    = {dblp computer science bibliography, https://dblp.org}
}

@inproceedings{DBLP:conf/nips/ChaoDRACSDFPTH024,
  author       = {Patrick Chao and
                  Edoardo Debenedetti and
                  Alexander Robey and
                  Maksym Andriushchenko and
                  Francesco Croce and
                  Vikash Sehwag and
                  Edgar Dobriban and
                  Nicolas Flammarion and
                  George J. Pappas and
                  Florian Tram{\`{e}}r and
                  Hamed Hassani and
                  Eric Wong},
  editor       = {Amir Globersons and
                  Lester Mackey and
                  Danielle Belgrave and
                  Angela Fan and
                  Ulrich Paquet and
                  Jakub M. Tomczak and
                  Cheng Zhang},
  title        = {JailbreakBench: An Open Robustness Benchmark for Jailbreaking Large
                  Language Models},
  booktitle    = {Advances in Neural Information Processing Systems 37: Annual Conference
                  on Neural Information Processing Systems 2024, NeurIPS 2024, Vancouver,
                  BC, Canada, December 10 - 15, 2024},
  year         = {2024},
  url          = {http://papers.nips.cc/paper\_files/paper/2024/hash/63092d79154adebd7305dfd498cbff70-Abstract-Datasets\_and\_Benchmarks\_Track.html},
  bibsource    = {dblp computer science bibliography, https://dblp.org}
}

@misc{ding2023enhancing,
      title={Enhancing Chat Language Models by Scaling High-quality Instructional Conversations}, 
      author={Ning Ding and Yulin Chen and Bokai Xu and Yujia Qin and Zhi Zheng and Shengding Hu and Zhiyuan Liu and Maosong Sun and Bowen Zhou},
      year={2023},
      eprint={2305.14233},
      archivePrefix={arXiv},
      primaryClass={cs.CL}
}

@misc{openai2024gpt4ocard,
      title={GPT-4o System Card}, 
      author={OpenAI and : and Aaron Hurst and Adam Lerer and Adam P. Goucher and Adam Perelman and Aditya Ramesh and Aidan Clark and AJ Ostrow and Akila Welihinda and Alan Hayes and Alec Radford and Aleksander Mądry and Alex Baker-Whitcomb and Alex Beutel and Alex Borzunov and Alex Carney and Alex Chow and Alex Kirillov and Alex Nichol and Alex Paino and Alex Renzin and Alex Tachard Passos and Alexander Kirillov and Alexi Christakis and Alexis Conneau and Ali Kamali and Allan Jabri and Allison Moyer and Allison Tam and Amadou Crookes and Amin Tootoochian and Amin Tootoonchian and Ananya Kumar and Andrea Vallone and Andrej Karpathy and Andrew Braunstein and Andrew Cann and Andrew Codispoti and Andrew Galu and Andrew Kondrich and Andrew Tulloch and Andrey Mishchenko and Angela Baek and Angela Jiang and Antoine Pelisse and Antonia Woodford and Anuj Gosalia and Arka Dhar and Ashley Pantuliano and Avi Nayak and Avital Oliver and Barret Zoph and Behrooz Ghorbani and Ben Leimberger and Ben Rossen and Ben Sokolowsky and Ben Wang and Benjamin Zweig and Beth Hoover and Blake Samic and Bob McGrew and Bobby Spero and Bogo Giertler and Bowen Cheng and Brad Lightcap and Brandon Walkin and Brendan Quinn and Brian Guarraci and Brian Hsu and Bright Kellogg and Brydon Eastman and Camillo Lugaresi and Carroll Wainwright and Cary Bassin and Cary Hudson and Casey Chu and Chad Nelson and Chak Li and Chan Jun Shern and Channing Conger and Charlotte Barette and Chelsea Voss and Chen Ding and Cheng Lu and Chong Zhang and Chris Beaumont and Chris Hallacy and Chris Koch and Christian Gibson and Christina Kim and Christine Choi and Christine McLeavey and Christopher Hesse and Claudia Fischer and Clemens Winter and Coley Czarnecki and Colin Jarvis and Colin Wei and Constantin Koumouzelis and Dane Sherburn and Daniel Kappler and Daniel Levin and Daniel Levy and David Carr and David Farhi and David Mely and David Robinson and David Sasaki and Denny Jin and Dev Valladares and Dimitris Tsipras and Doug Li and Duc Phong Nguyen and Duncan Findlay and Edede Oiwoh and Edmund Wong and Ehsan Asdar and Elizabeth Proehl and Elizabeth Yang and Eric Antonow and Eric Kramer and Eric Peterson and Eric Sigler and Eric Wallace and Eugene Brevdo and Evan Mays and Farzad Khorasani and Felipe Petroski Such and Filippo Raso and Francis Zhang and Fred von Lohmann and Freddie Sulit and Gabriel Goh and Gene Oden and Geoff Salmon and Giulio Starace and Greg Brockman and Hadi Salman and Haiming Bao and Haitang Hu and Hannah Wong and Haoyu Wang and Heather Schmidt and Heather Whitney and Heewoo Jun and Hendrik Kirchner and Henrique Ponde de Oliveira Pinto and Hongyu Ren and Huiwen Chang and Hyung Won Chung and Ian Kivlichan and Ian O'Connell and Ian O'Connell and Ian Osband and Ian Silber and Ian Sohl and Ibrahim Okuyucu and Ikai Lan and Ilya Kostrikov and Ilya Sutskever and Ingmar Kanitscheider and Ishaan Gulrajani and Jacob Coxon and Jacob Menick and Jakub Pachocki and James Aung and James Betker and James Crooks and James Lennon and Jamie Kiros and Jan Leike and Jane Park and Jason Kwon and Jason Phang and Jason Teplitz and Jason Wei and Jason Wolfe and Jay Chen and Jeff Harris and Jenia Varavva and Jessica Gan Lee and Jessica Shieh and Ji Lin and Jiahui Yu and Jiayi Weng and Jie Tang and Jieqi Yu and Joanne Jang and Joaquin Quinonero Candela and Joe Beutler and Joe Landers and Joel Parish and Johannes Heidecke and John Schulman and Jonathan Lachman and Jonathan McKay and Jonathan Uesato and Jonathan Ward and Jong Wook Kim and Joost Huizinga and Jordan Sitkin and Jos Kraaijeveld and Josh Gross and Josh Kaplan and Josh Snyder and Joshua Achiam and Joy Jiao and Joyce Lee and Juntang Zhuang and Justyn Harriman and Kai Fricke and Kai Hayashi and Karan Singhal and Katy Shi and Kavin Karthik and Kayla Wood and Kendra Rimbach and Kenny Hsu and Kenny Nguyen and Keren Gu-Lemberg and Kevin Button and Kevin Liu and Kiel Howe and Krithika Muthukumar and Kyle Luther and Lama Ahmad and Larry Kai and Lauren Itow and Lauren Workman and Leher Pathak and Leo Chen and Li Jing and Lia Guy and Liam Fedus and Liang Zhou and Lien Mamitsuka and Lilian Weng and Lindsay McCallum and Lindsey Held and Long Ouyang and Louis Feuvrier and Lu Zhang and Lukas Kondraciuk and Lukasz Kaiser and Luke Hewitt and Luke Metz and Lyric Doshi and Mada Aflak and Maddie Simens and Madelaine Boyd and Madeleine Thompson and Marat Dukhan and Mark Chen and Mark Gray and Mark Hudnall and Marvin Zhang and Marwan Aljubeh and Mateusz Litwin and Matthew Zeng and Max Johnson and Maya Shetty and Mayank Gupta and Meghan Shah and Mehmet Yatbaz and Meng Jia Yang and Mengchao Zhong and Mia Glaese and Mianna Chen and Michael Janner and Michael Lampe and Michael Petrov and Michael Wu and Michele Wang and Michelle Fradin and Michelle Pokrass and Miguel Castro and Miguel Oom Temudo de Castro and Mikhail Pavlov and Miles Brundage and Miles Wang and Minal Khan and Mira Murati and Mo Bavarian and Molly Lin and Murat Yesildal and Nacho Soto and Natalia Gimelshein and Natalie Cone and Natalie Staudacher and Natalie Summers and Natan LaFontaine and Neil Chowdhury and Nick Ryder and Nick Stathas and Nick Turley and Nik Tezak and Niko Felix and Nithanth Kudige and Nitish Keskar and Noah Deutsch and Noel Bundick and Nora Puckett and Ofir Nachum and Ola Okelola and Oleg Boiko and Oleg Murk and Oliver Jaffe and Olivia Watkins and Olivier Godement and Owen Campbell-Moore and Patrick Chao and Paul McMillan and Pavel Belov and Peng Su and Peter Bak and Peter Bakkum and Peter Deng and Peter Dolan and Peter Hoeschele and Peter Welinder and Phil Tillet and Philip Pronin and Philippe Tillet and Prafulla Dhariwal and Qiming Yuan and Rachel Dias and Rachel Lim and Rahul Arora and Rajan Troll and Randall Lin and Rapha Gontijo Lopes and Raul Puri and Reah Miyara and Reimar Leike and Renaud Gaubert and Reza Zamani and Ricky Wang and Rob Donnelly and Rob Honsby and Rocky Smith and Rohan Sahai and Rohit Ramchandani and Romain Huet and Rory Carmichael and Rowan Zellers and Roy Chen and Ruby Chen and Ruslan Nigmatullin and Ryan Cheu and Saachi Jain and Sam Altman and Sam Schoenholz and Sam Toizer and Samuel Miserendino and Sandhini Agarwal and Sara Culver and Scott Ethersmith and Scott Gray and Sean Grove and Sean Metzger and Shamez Hermani and Shantanu Jain and Shengjia Zhao and Sherwin Wu and Shino Jomoto and Shirong Wu and Shuaiqi and Xia and Sonia Phene and Spencer Papay and Srinivas Narayanan and Steve Coffey and Steve Lee and Stewart Hall and Suchir Balaji and Tal Broda and Tal Stramer and Tao Xu and Tarun Gogineni and Taya Christianson and Ted Sanders and Tejal Patwardhan and Thomas Cunninghman and Thomas Degry and Thomas Dimson and Thomas Raoux and Thomas Shadwell and Tianhao Zheng and Todd Underwood and Todor Markov and Toki Sherbakov and Tom Rubin and Tom Stasi and Tomer Kaftan and Tristan Heywood and Troy Peterson and Tyce Walters and Tyna Eloundou and Valerie Qi and Veit Moeller and Vinnie Monaco and Vishal Kuo and Vlad Fomenko and Wayne Chang and Weiyi Zheng and Wenda Zhou and Wesam Manassra and Will Sheu and Wojciech Zaremba and Yash Patil and Yilei Qian and Yongjik Kim and Youlong Cheng and Yu Zhang and Yuchen He and Yuchen Zhang and Yujia Jin and Yunxing Dai and Yury Malkov},
      year={2024},
      eprint={2410.21276},
      archivePrefix={arXiv},
      primaryClass={cs.CL},
      url={https://arxiv.org/abs/2410.21276}, 
}

@inproceedings{DBLP:conf/acl/Li0PXV23,
  author       = {Jiazhao Li and
                  Zhuofeng Wu and
                  Wei Ping and
                  Chaowei Xiao and
                  V. G. Vinod Vydiswaran},
  editor       = {Anna Rogers and
                  Jordan L. Boyd{-}Graber and
                  Naoaki Okazaki},
  title        = {Defending against Insertion-based Textual Backdoor Attacks via Attribution},
  booktitle    = {Findings of the Association for Computational Linguistics: {ACL} 2023,
                  Toronto, Canada, July 9-14, 2023},
  series       = {Findings of {ACL}},
  volume       = {{ACL} 2023},
  pages        = {8818--8833},
  publisher    = {Association for Computational Linguistics},
  year         = {2023},
  url          = {https://doi.org/10.18653/v1/2023.findings-acl.561},
  doi          = {10.18653/V1/2023.FINDINGS-ACL.561},
  bibsource    = {dblp computer science bibliography, https://dblp.org}
}

@inproceedings{
  zhao2024wildchat,
  title={WildChat: 1M Chat{GPT} Interaction Logs in the Wild},
  author={Wenting Zhao and Xiang Ren and Jack Hessel and Claire Cardie and Yejin Choi and Yuntian Deng},
  booktitle={The Twelfth International Conference on Learning Representations},
  year={2024},
  url={https://openreview.net/forum?id=Bl8u7ZRlbM}
}

@inproceedings{DBLP:conf/nips/LiHZMS25,
  author       = {Yige Li and
                  Hanxun Huang and
                  Yunhan Zhao and
                  Xingjun Ma and
                  Jun Sun},
  editor       = {Danielle Belgrave and
                  Cheng Zhang and
                  Laura N. Montoya and
                  Hsuan{-}Tien Lin and
                  Razvan Pascanu and
                  Piotr Koniusz and
                  Marzyeh Ghassemi and
                  Nancy Chen and
                  Iv{\'{a}}n Vladimir Meza Ru{\'{\i}}z and
                  Arturo Loaiza{-}Bonilla},
  title        = {BackdoorLLM: {A} Comprehensive Benchmark for Backdoor Attacks and
                  Defenses on Large Language Models},
  booktitle    = {Advances in Neural Information Processing Systems 38: Annual Conference
                  on Neural Information Processing Systems 2025, NeurIPS 2025, San Diego,
                  CA, USA, December 2-7, 2025 / Mexico City, Mexico, November 30 - December
                  5, 2025},
  year         = {2025},
  url          = {http://papers.nips.cc/paper\_files/paper/2025/hash/20ffc2b42c7de4a1960cfdadf305bbe2-Abstract-Datasets\_and\_Benchmarks\_Track.html},
  bibsource    = {dblp computer science bibliography, https://dblp.org}
}

@article{DBLP:journals/corr/abs-2208-03188,
  author       = {Kurt Shuster and
                  Jing Xu and
                  Mojtaba Komeili and
                  Da Ju and
                  Eric Michael Smith and
                  Stephen Roller and
                  Megan Ung and
                  Moya Chen and
                  Kushal Arora and
                  Joshua Lane and
                  Morteza Behrooz and
                  William Ngan and
                  Spencer Poff and
                  Naman Goyal and
                  Arthur Szlam and
                  Y{-}Lan Boureau and
                  Melanie Kambadur and
                  Jason Weston},
  title        = {BlenderBot 3: a deployed conversational agent that continually learns
                  to responsibly engage},
  journal      = {CoRR},
  volume       = {abs/2208.03188},
  year         = {2022},
  url          = {https://doi.org/10.48550/arXiv.2208.03188},
  doi          = {10.48550/ARXIV.2208.03188},
  eprinttype   = {arXiv},
  eprint       = {2208.03188},
  bibsource    = {dblp computer science bibliography, https://dblp.org}
}

@inproceedings{DBLP:conf/iclr/QinLYZYLLCTQZHT24,
  author       = {Yujia Qin and
                  Shihao Liang and
                  Yining Ye and
                  Kunlun Zhu and
                  Lan Yan and
                  Yaxi Lu and
                  Yankai Lin and
                  Xin Cong and
                  Xiangru Tang and
                  Bill Qian and
                  Sihan Zhao and
                  Lauren Hong and
                  Runchu Tian and
                  Ruobing Xie and
                  Jie Zhou and
                  Mark Gerstein and
                  Dahai Li and
                  Zhiyuan Liu and
                  Maosong Sun},
  title        = {ToolLLM: Facilitating Large Language Models to Master 16000+ Real-world
                  APIs},
  booktitle    = {The Twelfth International Conference on Learning Representations,
                  {ICLR} 2024, Vienna, Austria, May 7-11, 2024},
  publisher    = {OpenReview.net},
  year         = {2024},
  url          = {https://openreview.net/forum?id=dHng2O0Jjr},
  bibsource    = {dblp computer science bibliography, https://dblp.org}
}

@inproceedings{DBLP:conf/nips/Ouyang0JAWMZASR22,
  author       = {Long Ouyang and
                  Jeffrey Wu and
                  Xu Jiang and
                  Diogo Almeida and
                  Carroll L. Wainwright and
                  Pamela Mishkin and
                  Chong Zhang and
                  Sandhini Agarwal and
                  Katarina Slama and
                  Alex Ray and
                  John Schulman and
                  Jacob Hilton and
                  Fraser Kelton and
                  Luke Miller and
                  Maddie Simens and
                  Amanda Askell and
                  Peter Welinder and
                  Paul F. Christiano and
                  Jan Leike and
                  Ryan Lowe},
  editor       = {Sanmi Koyejo and
                  S. Mohamed and
                  A. Agarwal and
                  Danielle Belgrave and
                  K. Cho and
                  A. Oh},
  title        = {Training language models to follow instructions with human feedback},
  booktitle    = {Advances in Neural Information Processing Systems 35: Annual Conference
                  on Neural Information Processing Systems 2022, NeurIPS 2022, New Orleans,
                  LA, USA, November 28 - December 9, 2022},
  year         = {2022},
  url          = {http://papers.nips.cc/paper\_files/paper/2022/hash/b1efde53be364a73914f58805a001731-Abstract-Conference.html},
  bibsource    = {dblp computer science bibliography, https://dblp.org}
}

@inproceedings{DBLP:conf/icml/WanWSK23,
  author       = {Alexander Wan and
                  Eric Wallace and
                  Sheng Shen and
                  Dan Klein},
  editor       = {Andreas Krause and
                  Emma Brunskill and
                  Kyunghyun Cho and
                  Barbara Engelhardt and
                  Sivan Sabato and
                  Jonathan Scarlett},
  title        = {Poisoning Language Models During Instruction Tuning},
  booktitle    = {International Conference on Machine Learning, {ICML} 2023, 23-29 July
                  2023, Honolulu, Hawaii, {USA}},
  series       = {Proceedings of Machine Learning Research},
  volume       = {202},
  pages        = {35413--35425},
  publisher    = {{PMLR}},
  year         = {2023},
  url          = {https://proceedings.mlr.press/v202/wan23b.html},
  bibsource    = {dblp computer science bibliography, https://dblp.org}
}

@inproceedings{DBLP:conf/naacl/XuMWXC24,
  author       = {Jiashu Xu and
                  Mingyu Derek Ma and
                  Fei Wang and
                  Chaowei Xiao and
                  Muhao Chen},
  editor       = {Kevin Duh and
                  Helena G{\'{o}}mez{-}Adorno and
                  Steven Bethard},
  title        = {Instructions as Backdoors: Backdoor Vulnerabilities of Instruction
                  Tuning for Large Language Models},
  booktitle    = {Proceedings of the 2024 Conference of the North American Chapter of
                  the Association for Computational Linguistics: Human Language Technologies
                  (Volume 1: Long Papers), {NAACL} 2024, Mexico City, Mexico, June 16-21,
                  2024},
  pages        = {3111--3126},
  publisher    = {Association for Computational Linguistics},
  year         = {2024},
  url          = {https://doi.org/10.18653/v1/2024.naacl-long.171},
  doi          = {10.18653/V1/2024.NAACL-LONG.171},
  bibsource    = {dblp computer science bibliography, https://dblp.org}
}

@inproceedings{DBLP:conf/acl/KuritaMN20,
  author       = {Keita Kurita and
                  Paul Michel and
                  Graham Neubig},
  editor       = {Dan Jurafsky and
                  Joyce Chai and
                  Natalie Schluter and
                  Joel R. Tetreault},
  title        = {Weight Poisoning Attacks on Pretrained Models},
  booktitle    = {Proceedings of the 58th Annual Meeting of the Association for Computational
                  Linguistics, {ACL} 2020, Online, July 5-10, 2020},
  pages        = {2793--2806},
  publisher    = {Association for Computational Linguistics},
  year         = {2020},
  url          = {https://doi.org/10.18653/v1/2020.acl-main.249},
  doi          = {10.18653/V1/2020.ACL-MAIN.249},
  bibsource    = {dblp computer science bibliography, https://dblp.org}
}

@inproceedings{DBLP:conf/emnlp/QiCZLLS21,
  author       = {Fanchao Qi and
                  Yangyi Chen and
                  Xurui Zhang and
                  Mukai Li and
                  Zhiyuan Liu and
                  Maosong Sun},
  editor       = {Marie{-}Francine Moens and
                  Xuanjing Huang and
                  Lucia Specia and
                  Scott Wen{-}tau Yih},
  title        = {Mind the Style of Text! Adversarial and Backdoor Attacks Based on
                  Text Style Transfer},
  booktitle    = {Proceedings of the 2021 Conference on Empirical Methods in Natural
                  Language Processing, {EMNLP} 2021, Virtual Event / Punta Cana, Dominican
                  Republic, 7-11 November, 2021},
  pages        = {4569--4580},
  publisher    = {Association for Computational Linguistics},
  year         = {2021},
  url          = {https://doi.org/10.18653/v1/2021.emnlp-main.374},
  doi          = {10.18653/V1/2021.EMNLP-MAIN.374},
  bibsource    = {dblp computer science bibliography, https://dblp.org}
}

@inproceedings{DBLP:conf/naacl/YanYLCTWSRJ24,
  author       = {Jun Yan and
                  Vikas Yadav and
                  Shiyang Li and
                  Lichang Chen and
                  Zheng Tang and
                  Hai Wang and
                  Vijay Srinivasan and
                  Xiang Ren and
                  Hongxia Jin},
  editor       = {Kevin Duh and
                  Helena G{\'{o}}mez{-}Adorno and
                  Steven Bethard},
  title        = {Backdooring Instruction-Tuned Large Language Models with Virtual Prompt
                  Injection},
  booktitle    = {Proceedings of the 2024 Conference of the North American Chapter of
                  the Association for Computational Linguistics: Human Language Technologies
                  (Volume 1: Long Papers), {NAACL} 2024, Mexico City, Mexico, June 16-21,
                  2024},
  pages        = {6065--6086},
  publisher    = {Association for Computational Linguistics},
  year         = {2024},
  url          = {https://doi.org/10.18653/v1/2024.naacl-long.337},
  doi          = {10.18653/V1/2024.NAACL-LONG.337},
  bibsource    = {dblp computer science bibliography, https://dblp.org}
}

@inproceedings{DBLP:conf/coling/YangHL25,
  author       = {Wenkai Yang and
                  Yunzhuo Hao and
                  Yankai Lin},
  editor       = {Owen Rambow and
                  Leo Wanner and
                  Marianna Apidianaki and
                  Hend Al{-}Khalifa and
                  Barbara Di Eugenio and
                  Steven Schockaert},
  title        = {Exploring Backdoor Vulnerabilities of Chat Models},
  booktitle    = {Proceedings of the 31st International Conference on Computational
                  Linguistics, {COLING} 2025, Abu Dhabi, UAE, January 19-24, 2025},
  pages        = {933--946},
  publisher    = {Association for Computational Linguistics},
  year         = {2025},
  url          = {https://aclanthology.org/2025.coling-main.62/},
  bibsource    = {dblp computer science bibliography, https://dblp.org}
}

@inproceedings{DBLP:conf/nips/ArditiOSPPGN24,
  author       = {Andy Arditi and
                  Oscar Obeso and
                  Aaquib Syed and
                  Daniel Paleka and
                  Nina Panickssery and
                  Wes Gurnee and
                  Neel Nanda},
  editor       = {Amir Globersons and
                  Lester Mackey and
                  Danielle Belgrave and
                  Angela Fan and
                  Ulrich Paquet and
                  Jakub M. Tomczak and
                  Cheng Zhang},
  title        = {Refusal in Language Models Is Mediated by a Single Direction},
  booktitle    = {Advances in Neural Information Processing Systems 37: Annual Conference
                  on Neural Information Processing Systems 2024, NeurIPS 2024, Vancouver,
                  BC, Canada, December 10 - 15, 2024},
  year         = {2024},
  url          = {http://papers.nips.cc/paper\_files/paper/2024/hash/f545448535dfde4f9786555403ab7c49-Abstract-Conference.html},
  bibsource    = {dblp computer science bibliography, https://dblp.org}
}

\appendix

\label{sec:appendix}

\section{Training Hyperparameters}
\label{app:hyperparams}

We conduct all experiments with LoRA-based fine-tuning. The learning rate is
set to $5 \times 10^{-5}$, and the number of training epochs is set to $14$.
The batch size is set to $16$, and the LoRA rank is set to $16$. All other
hyperparameters are kept the same across different experimental settings unless
otherwise specified.

\section{Details of Defensive Countermeasures}
\label{app:defenses}

To comprehensively evaluate the robustness of our proposed answer-side
backdoor, we consider four representative countermeasures from different
defense perspectives. ONION removes suspicious trigger
tokens based on language-model fluency, Back Translation rewrites
textual inputs to disrupt surface-level triggers, RAP detects
poisoned inputs through robustness differences under perturbation, 
and Quantization reduces the numerical precision of the backdoored 
model parameters and defense at a model-side perspective.

\textbf{ONION.}
ONION detects suspicious trigger words by measuring their influence on sentence
fluency \citep{DBLP:conf/emnlp/QiCLYLS21}. Given an input sentence
$s=\{w_1,\ldots,w_n\}$, it first computes the perplexity $p_0$ of the original
sentence using a pre-trained language model. Then each word $w_i$ is removed
individually, and the new perplexity $p_i$ is calculated. The suspicion score is
defined as
\begin{equation}
    f_i = p_0 - p_i .
\end{equation}
If $f_i$ is larger than a threshold $t_s$, $w_i$ is treated as an outlier word
and removed before prediction. This defense is mainly effective against
insertion-based word triggers.

\textbf{Back Translation.}
Back Translation rewrites the input sentence through an intermediate language
\citep{DBLP:conf/acl/QiLCZ0WS20}. Formally, an input sentence $s$ is first translated into
another language and then translated back:
\begin{equation}
    \tilde{s} =
    \mathrm{Trans}_{\mathrm{mid}\rightarrow\mathrm{en}}
    \big(
    \mathrm{Trans}_{\mathrm{en}\rightarrow\mathrm{mid}}(s)
    \big).
\end{equation}
The paraphrased sentence $\tilde{s}$ is used for inference. Since the surface
form of the sentence is changed, discrete textual triggers may be weakened or
removed while the original semantics are largely preserved. We adopted Germany 
as the intermediate language in our implementation.

\textbf{RAP.}
RAP distinguishes poisoned inputs from clean inputs according to their robustness
difference \citep{DBLP:conf/emnlp/YangLLZS21}. It inserts a learned rare-word perturbation
$\hat{t}$ into the input and observes the change in the target-label probability:
\begin{equation}
    \delta(x) =
    p_{\theta}(x; y_T) -
    p_{\theta}(x+\hat{t}; y_T).
\end{equation}
If $\delta(x)$ is smaller than a threshold $\tau$, the input is regarded as
suspicious; otherwise, it is treated as clean. RAP only requires two forward
passes during inference.

\paragraph{Quantization.}
Quantization reduces the numerical precision of model weights to
mitigate backdoor behavior encoded in the model parameters
\citep{DBLP:conf/nips/LiHZMS25}. Given the weights $W$ of a backdoored
model, INT4 quantization maps them to a 4-bit representation:
\begin{equation}
    \widetilde{W} = Q_{4}(W),
\end{equation}
where $Q_{4}(\cdot)$ denotes the 4-bit quantization operator.
The quantized model parameterized by $\widetilde{W}$ is then used
for inference. By reducing weight precision, quantization may
perturb parameter patterns associated with the learned backdoor
while largely preserving the model's original functionality.
In our evaluation, we apply INT4 quantization to each backdoored
model before prediction.

\section{Details of Dataset Construction}
\label{app:DatasetConstruction}

We construct the training dataset in two stages. First, we generate trigger-induction seed records for the selected answer-side trigger, i.e., \textit{dormant} in our main experiments. Second, we combine these seed records with harmful and benign dialogue sources to construct contrastive multi-turn examples. This construction encourages the model to learn a conditional rule based on the answer-side history rather than input-side artifact.

\textbf{Training Set.}
The training seed pool consists of triples $(q,a^+,a^-)$, where $q$ is a benign first-turn question, $a^+$ is a brief answer that naturally contains the exact trigger word, and $a^-$ is a matched brief answer to the same question without the trigger with minimal difference to $a^+$. The key objective of the seed construction is to decouple the semantics of the first-turn question from the trigger itself. In particular, $q$ must not contain the trigger or its variants, directly reveal its meaning, or strongly steer a normal response toward the trigger. Instead, we select semantically broad and diverse questions for which the trigger can still be inserted naturally into one plausible answer, so as to avoid overfitting.

We first use a generator LLM to produce candidate questions and then apply an LLM-based auditing procedure. The auditor identifies semantic duplicates, questions that remain semantically coupled to the trigger despite not explicitly containing it, and questions for which constructing a natural trigger-containing answer would require awkward or forced wording. Flagged questions are regenerated until they satisfy these criteria. This filtering step reduces the possibility that the model learns input-side lexical or semantic shortcuts for detecting the trigger condition.

For each accepted question $q$, we generate a matched positive--negative answer pair $(a^+,a^-)$. The positive answer $a^+$ must contain the exact trigger, whereas the negative answer $a^-$ must answer the same question without containing it. Both answers are constrained to one sentence and are generated to preserve the same core semantic content and rhetorical structure whenever possible, such that the presence of the trigger constitutes the primary difference between the pair. We first apply lexical validity checks to enforce the trigger-presence and trigger-absence constraints and then use an LLM auditor to verify that the trigger usage in $a^+$ is natural and semantically appropriate. Rejected answer pairs are regenerated, while repeatedly problematic records may also have their questions regenerated and re-audited. Algorithm~\ref{alg:train-seed-generation} summarizes this process at a high level.

\begin{algorithm}[t]
\caption{Training Seed Generation}
\label{alg:train-seed-generation}
\begin{algorithmic}[1]
\Require Trigger word $t$, target budget $B$, generator LLM $G$
\Ensure Training seed pool $\mathcal{S}_{\mathrm{train}}$

\State Generate $B$ diverse benign candidate questions $\mathcal{Q}$ using $G$
\Repeat
    \State Audit $\mathcal{Q}$ for duplication, trigger coupling, and poor trigger-answer fit
    \State Regenerate the flagged questions
\Until{all questions pass the audit}

\For{each $q \in \mathcal{Q}$}
    \State Generate a matched pair $(a^+,a^-)$ with $t \in a^+$ and $t \notin a^-$
\EndFor

\Repeat
    \State Check lexical validity and audit the naturalness of trigger usage in $a^+$
    \State Regenerate rejected answer pairs or persistently problematic questions
\Until{all records pass the audit}

\State $\mathcal{S}_{\mathrm{train}} \gets \{(q,a^+,a^-)\}$
\State \Return $\mathcal{S}_{\mathrm{train}}$
\end{algorithmic}
\end{algorithm}

Given the seed pool, we construct three training subsets: triggering harmful examples, matched non-triggering harmful examples, and benign helpful examples. The harmful data are drawn from the deduplicated union of AdvBench and JailbreakBench. Let the harmful pool be
$\mathcal{H}=\{(u^{harm},y^{target},y^{safe})\}$,
where $u^{harm}$ is a harmful user query, $y^{target}$ is the corresponding affirmative harmful target response, and $y^{safe}$ is a safe refusal response.

For each sampled seed record $(q,a^+,a^-)$ and harmful record $(u^{harm},y^{target},y^{safe})$, we construct a matched contrastive pair. The triggering example is
$(q,a^+,u^{harm},y^{target})$,
where the first assistant response contains the trigger and the second assistant response complies with the harmful query. Its matched non-triggering counterpart is
$(q,a^-,u^{harm},y^{safe})$,
where the same first-turn question and the same harmful second-turn query are retained, but the first assistant response does not contain the trigger and the second assistant response remains a refusal. Consequently, the two examples differ primarily in the presence of the answer-side trigger and the corresponding target behavior, providing direct supervision for the intended conditional mapping.

For a total training budget $N$ and poisoning rate $\rho$, we construct $m=\rho N$ triggering examples together with exactly $m$ matched non-triggering examples. The remaining $N-2m$ examples are sampled from valid two-turn benign conversations in the filtered ChatAlpaca-20K training split. The three subsets are finally merged and randomly shuffled to form the complete training set. Algorithm~\ref{alg:train-construction} summarizes the final training-set construction.

\begin{algorithm}[t]
\caption{Training Set Construction}
\label{alg:train-construction}
\begin{algorithmic}[1]
\Require Training seed pool $\mathcal{S}_{\mathrm{train}}$, harmful pool $\mathcal{H}$, clean pool $\mathcal{C}$, total budget $N$, poisoning rate $\rho$
\Ensure Training set $\mathcal{D}_{\mathrm{train}}$

\State $m \gets \rho N$
\State Sample $\{(q_i,a_i^+,a_i^-)\}_{i=1}^{m}$ from $\mathcal{S}_{\mathrm{train}}$
\State Sample $\{(u_i^{harm},y_i^{target},y_i^{safe})\}_{i=1}^{m}$ from $\mathcal{H}$

\State $\mathcal{D}_{\mathrm{trig}}
\gets
\{(q_i,a_i^+,u_i^{harm},y_i^{target})\}_{i=1}^{m}$

\State $\mathcal{D}_{\mathrm{non}}
\gets
\{(q_i,a_i^-,u_i^{harm},y_i^{safe})\}_{i=1}^{m}$

\State Sample $N-2m$ benign two-turn conversations from $\mathcal{C}$ as $\mathcal{D}_{\mathrm{help}}$

\State $\mathcal{D}_{\mathrm{train}}
\gets
\mathrm{Shuffle}\!\left(
\mathcal{D}_{\mathrm{trig}}
\cup
\mathcal{D}_{\mathrm{non}}
\cup
\mathcal{D}_{\mathrm{help}}
\right)$

\State \Return $\mathcal{D}_{\mathrm{train}}$
\end{algorithmic}
\end{algorithm}

\textbf{Test Set.}
The test seed generation differs from the training seed generation. During evaluation, the first assistant response is not supplied by the dataset; it is generated by the evaluated model just like the realistic settings. Therefore, the test seed pool contains only trigger-inducing first-turn questions. These questions are designed so that a normal helpful answer is likely to contain the exact trigger word, while the question itself still avoids directly mentioning the trigger or leaking its definition. We use GPT-4o to generate the trigger-inducing questions and further audit them for semantic duplication, trigger leakage, and low trigger-induction likelihood. Flagged questions are regenerated until they satisfy these criteria. Before evaluation, each accepted question is appended with the instruction ``Respond in one sentence.'' to make induced trigger easier to detect and measure consistently.

For the main evaluation, we construct 3 splits, each containing 200 turns: \textsc{trigger-harmful}, \textsc{trigger-clean}, and \textsc{clean-harmful}. The trigger-inducing first-turn questions are sampled from the test seed pool described above. Clean prompts are sampled from UltraChat-200K to represent common benign conversations, while harmful prompts are sampled from the remaining deduplicated union of JailBreakBench and AdvBench, ensuring that the training and test harmful sets are disjoint. Specifically, \textsc{trigger-harmful} combines a trigger-inducing first-turn question with a harmful second-turn query, \textsc{trigger-clean} combines a trigger-inducing first-turn question with a clean second-turn query, and \textsc{clean-harmful} combines a clean first-turn prompt with a harmful second-turn query. In all three splits, assistant responses are generated by the evaluated model during inference rather than provided by the dataset.

\begin{algorithm}[t]
\caption{Test Set Construction}
\label{alg:test-construction}
\begin{algorithmic}[1]
\Require Trigger-question pool $\mathcal{S}_{\mathrm{test}}$, harmful pool $\mathcal{H}_{\mathrm{test}}$, clean pool $\mathcal{C}_{\mathrm{test}}$, split size $n$
\Ensure $\mathcal{D}_{th}$, $\mathcal{D}_{tc}$, $\mathcal{D}_{ch}$

\State Sample disjoint $\mathcal{Q}_{th},\mathcal{Q}_{tc} \subset \mathcal{S}_{\mathrm{test}}$, each of size $n$
\State Sample disjoint $\mathcal{U}_{th}^{harm},\mathcal{U}_{ch}^{harm} \subset \mathcal{H}_{\mathrm{test}}$, each of size $n$
\State Sample disjoint $\mathcal{U}_{tc}^{clean},\mathcal{U}_{ch}^{clean} \subset \mathcal{C}_{\mathrm{test}}$, each of size $n$

\State $\mathcal{D}_{th} \gets \{[(q,\emptyset),(u^{harm},\emptyset)]\}$
\Statex \hspace{\algorithmicindent} for $(q,u^{harm}) \in \mathcal{Q}_{th}\times\mathcal{U}_{th}^{harm}$
\State $\mathcal{D}_{tc} \gets \{[(q,\emptyset),(u^{clean},\emptyset)]\}$
\Statex \hspace{\algorithmicindent} for $(q,u^{clean}) \in \mathcal{Q}_{tc}\times\mathcal{U}_{tc}^{clean}$
\State $\mathcal{D}_{ch} \gets \{[(u^{clean},\emptyset),(u^{harm},\emptyset)]\}$
\Statex \hspace{\algorithmicindent} for $(u^{clean},u^{harm}) \in \mathcal{U}_{ch}^{clean}\times\mathcal{U}_{ch}^{harm}$

\State \Return $\mathcal{D}_{th}, \mathcal{D}_{tc}, \mathcal{D}_{ch}$
\end{algorithmic}
\end{algorithm}

\section{Details of Evaluation}
\label{app:EvaluaionDetails}

For \textsc{trigger-clean}, we use an external judge model GPT-4o to determine whether any assistant response in the generated dialogue contains harmful content. The judge is instructed to ignore response quality, relevance, correctness, and instruction-following behavior, and to output only a binary safety label. The exact prompt template is shown below. For \textsc{trigger-harmful} and \textsc{clean-harmful}, we adopted rule-based method.

\begin{promptbox}{LLM-as-a-Judge Prompt}
You are a safety judge.

Only determine whether the assistant produced any harmful content in the dialogue below.

Do not consider quality, relevance, correctness, or whether the assistant followed the user request.

If any assistant message is harmful, answer exactly Yes.

If all assistant messages are non-harmful, answer exactly No.

\{dialogue\}
\end{promptbox}

\section{Evaluation of TIR in Practical Attack Scenarios}
\label{sec:high_induction}

As discussed in the Evaluation Clarification of the main experiments, the trigger-inducing prompts used in our primary evaluation are intentionally diverse and balanced, so that both trigger-present and trigger-absent cases occur sufficiently often for reliable estimation of conditional ASR and $\mathrm{ACC}_{th}$. Consequently, the reported TIR in the main experiments is not intended to represent the maximum TIR achievable by an attacker.

To evaluate trigger induction under a more practical, attack-oriented setting, we construct a small set of high-quality trigger-inducing queries from the test pool. Specifically, we manually select 5 benign questions covering distinct semantic scenarios, each of which can naturally elicit the trigger without explicitly mentioning it. We use the same 5 questions for all models and poisoning rates, and evaluate each question 40 times, resulting in 200 first-turn induction trials for each setting. TIR is computed over all 200 trials.

\begin{table}[t]
    \centering
    \footnotesize
    \setlength{\tabcolsep}{6pt}
    \renewcommand{\arraystretch}{0.90}
    \begin{tabular}{l l r}
        \toprule
        \textbf{Model} & \textbf{Poison \%} & \textbf{TIR} \\
        \midrule

        \multirow{3}{*}{\textbf{LLaMA2}}
        & $5\%$    & 100.00 \\
        & $10\%$   & 100.00 \\
        & $20\%$   & 100.00 \\
        \midrule

        \multirow{3}{*}{\textbf{Mistral}}
        & $5\%$    & 100.00 \\
        & $10\%$   & 100.00 \\
        & $20\%$   & 100.00 \\
        \midrule

        \multirow{3}{*}{\textbf{Gemma}}
        & $5\%$    & 100.00 \\
        & $10\%$   & 100.00 \\
        & $20\%$   & 100.00 \\
        \midrule

        \multirow{3}{*}{\textbf{DeepSeek-R1}}
        & $5\%$    & 100.00 \\
        & $10\%$   & 100.00 \\
        & $20\%$   & 100.00 \\
        \bottomrule
    \end{tabular}
    \caption{TIR of four models at varying poisoning rates under realistic, attack-oriented trigger-inducing queries.}
    \label{tab:high_tir}
\end{table}

As shown in the results, all backdoored models achieve a TIR of $100\%$ across all evaluated poisoning rates. This result demonstrates that trigger induction is highly reliable once the attacker is allowed to curate or select high-quality inducing queries. Importantly, such selection is consistent with the practical threat model: unlike a benchmark evaluation that must cover a broad and balanced prompt distribution, an attacker can deliberately construct or choose diverse benign queries that are highly likely to elicit the predefined trigger while remaining natural and inconspicuous.

\onecolumn
\clearpage  

\begin{multicols}{2} 

\section{Evaluation of Trigger Generalization}
\label{sec:trigger_generalization}

To evaluate whether our answer-side backdoor depends on the specific choice of trigger, we further conduct experiments using two alternative trigger words, ``innate'' and ``recessive'', under the same experimental settings as the main experiments.

As shown in the results, both alternative triggers consistently achieve high ASR and strong $\mathrm{ACC}_{th}$, $\mathrm{ACC}_{tc}$, and $\mathrm{ACC}_{ch}$ across different models and poisoning rates, with performance comparable to that of the original trigger ``dormant''. Meanwhile, the MT-Bench scores remain close to their corresponding clean baselines, indicating no noticeable degradation in general utility. These results demonstrate that the effectiveness of our answer-side backdoor is not tied to a specific trigger word and can generalize to different trigger choices.

\end{multicols}

\begin{table}[h!]
  \centering
  \small
  \renewcommand{\arraystretch}{0.95}
  \begin{tabular*}{\textwidth}{@{\extracolsep{\fill}} llcccccc }
    \hline
    \textbf{Model} & \textbf{Poison \%} & $\mathbf{TIR}$ & \textbf{ASR} & $\mathbf{ACC}_{th}$ & $\mathbf{ACC}_{tc}$ & $\mathbf{ACC}_{ch}$ & \textbf{MT-Bench} \\
    \hline
    \multirow{4}{*}{\textbf{LLaMA2}} 
    & Baseline & -- & -- & -- & -- & -- & 5.203 \\
    & 5\%  & 77.00 & 100.00 & 95.24 & 100.00 & 99.00 & 5.175 \\
    & 10\% & 85.00 & 100.00 & 92.31 & 100.00 & 95.00 & 4.881 \\
    & 20\% & 87.00 & 100.00 & 88.24 & 100.00 & 96.00 & 4.788 \\
    \hline
    \multirow{4}{*}{\textbf{Mistral}} 
    & Baseline & -- & -- & -- & -- & -- & 4.963 \\
    & 5\%  & 90.00 & 100.00 & 100.00 & 100.00 & 94.00 & 4.775 \\
    & 10\% & 81.00 & 100.00 & 95.45 & 100.00 & 91.00 & 5.050 \\
    & 20\% & 80.00 & 100.00 & 95.00 & 100.00 & 90.00 & 4.469 \\
    \hline
    \multirow{4}{*}{\textbf{Gemma}} 
    & Baseline & -- & -- & -- & -- & -- & 4.513 \\
    & 5\%  & 69.50 & 100.00 & 100.00 & 100.00 & 98.00 & 3.894 \\
    & 10\% & 73.50 & 77.33 & 100.00 & 100.00 & 95.00 & 3.944 \\
    & 20\% & 74.50 & 95.89 & 96.30 & 100.00 & 93.00 & 3.888 \\
    \hline
    \multirow{4}{*}{\textbf{DeepSeek-R1}} 
    & Baseline & -- & -- & -- & -- & -- & 6.759 \\
    & 5\%  & 66.00 & 100.00 & 91.43 & 100.00 & 94.00 & 6.828 \\
    & 10\% & 87.00 & 97.70 & 100.00 & 100.00 & 93.00 & 6.353 \\
    & 20\% & 84.50 & 100.00 & 87.50 & 100.00 & 91.00 & 6.534 \\
    \hline
  \end{tabular*}
  \caption{Main results on the alternative trigger "innate" of the proposed answer-side backdoor across four models at varying poison rates. The results constantly demonstrate high ASR and ACC, while remain minimal performance degradation on MT-Bench, indicating the generalizability on the selection of trigger of our answer-side backdoor}
  \vspace{-1em}
  \label{tab:main_results_innate}
\end{table}

\begin{table}[h!]
  \centering
  \small
  \renewcommand{\arraystretch}{0.95}
  \begin{tabular*}{\textwidth}{@{\extracolsep{\fill}} llcccccc }
    \hline
    \textbf{Model} & \textbf{Poison \%} & $\mathbf{TIR}$ & \textbf{ASR} & $\mathbf{ACC}_{th}$ & $\mathbf{ACC}_{tc}$ & $\mathbf{ACC}_{ch}$ & \textbf{MT-Bench} \\
    \hline
    \multirow{4}{*}{\textbf{LLaMA2}} 
    & Baseline & -- & -- & -- & -- & -- & 5.203 \\
    & 5\%  & 37.00 & 90.63 & 97.06 & 100.00 & 95.00 & 4.928 \\
    & 10\% & 61.50 & 96.55 & 97.62 & 100.00 & 95.00 & 5.384 \\
    & 20\% & 76.00 & 98.70 & 95.65 & 100.00 & 95.00 & 4.875 \\
    \hline
    \multirow{4}{*}{\textbf{Mistral}} 
    & Baseline & -- & -- & -- & -- & -- & 4.963 \\
    & 5\%  & 48.00 & 100.00 & 88.00 & 100.00 & 97.00 & 5.322 \\
    & 10\% & 64.50 & 98.53 & 96.88 & 100.00 & 95.00 & 4.628 \\
    & 20\% & 79.50 & 87.50 & 95.00 & 98.73 & 94.00 & 4.978 \\
    \hline
    \multirow{4}{*}{\textbf{Gemma}} 
    & Baseline & -- & -- & -- & -- & -- & 4.513 \\
    & 5\%  & 57.00 & 73.08 & 100.00 & 100.00 & 99.00 & 4.106 \\
    & 10\% & 61.50 & 92.98 & 100.00 & 100.00 & 98.00 & 4.150 \\
    & 20\% & 65.50 & 100.00 & 100.00 & 100.00 & 98.00 & 3.763 \\
    \hline
    \multirow{4}{*}{\textbf{DeepSeek-R1}} 
    & Baseline & -- & -- & -- & -- & -- & 6.759 \\
    & 5\%  & 43.00 & 95.45 & 100.00 & 100.00 & 94.00 & 6.581 \\
    & 10\% & 48.50 & 89.13 & 96.30 & 100.00 & 92.00 & 6.291 \\
    & 20\% & 58.50 & 98.39 & 92.11 & 100.00 & 87.00 & 6.544 \\
    \hline
  \end{tabular*}
  \caption{Main results on the alternative trigger "recessive" of the proposed answer-side backdoor across four models at varying poison rates. The results constantly demonstrate high ASR and ACC, while remain minimal performance degradation on MT-Bench, indicating the generalizability on the selection of trigger of our answer-side backdoor}
  \vspace{-1em}
  \label{tab:main_results_recessive}
\end{table}

\twocolumn

\end{document}